\documentclass[a4paper, amsfonts, amssymb, amsmath, reprint, showkeys, nofootinbib, twoside, preprintnumbers,aps,prd]{revtex4-2}
\usepackage{babel}
\usepackage{graphicx}
\usepackage{dcolumn}
\usepackage{mathrsfs}
\usepackage{bm}
\usepackage{xcolor}
\usepackage{mathtools}
\usepackage{amsthm}
\usepackage{verbatim}
\usepackage{hyperref}
\usepackage{empheq}
\hypersetup{colorlinks}

\begin{document}
\preprint{PI/UAN-2026-733FT}
\title{Jacobson’s thermodynamic approach to classical gravity applied to non-Riemannian geometries:  remarks on the simplicity of Nature}
\author{Jhan N. Martínez}
    \email{jhamarlo@correo.uis.edu.co}
    \affiliation{Escuela de F\'isica, Universidad Industrial de Santander,  Ciudad Universitaria, Bucaramanga 680002, Colombia}
\author{José F. Rodríguez-Ruiz }
    \thanks{A member of the ICRANet network (\url{https://www.icranet.org/})}
    \email{jrodriguez154@uan.edu.co}
    \affiliation{Departamento de F\'isica, Universidad Antonio Nari\~no, Cra 3 Este \# 47A - 15, Bogot\'a D.C. 110231, Colombia}
\author{Yeinzon Rodr\'iguez }
    \email{yeinzon@saber.uis.edu.co}
    \affiliation{ Centro de Investigaciones en Ciencias B\'asicas y Aplicadas, Universidad Antonio Nari\~no, Cra 3 Este \# 47A - 15, Bogot\'a D.C. 110231, Colombia\\}
    \affiliation{Escuela de F\'isica, Universidad Industrial de Santander,  Ciudad Universitaria, Bucaramanga 680002, Colombia\\}

\begin{abstract}
Three decades ago, Ted Jacobson surprised us with a very appealing approach to 
classical gravity.  According to him, the gravitational field equations are the consequence of the first law of thermodynamics applied to a Rindler observer.    
Jacobson’s approach being formulated for Riemannian geometries, we have wondered what its consequences would be for non-Riemannian geometries.  The results of our quest have been particularly appealing:  we have found that the theory that derives from the Einstein-Hilbert action, arguably ``the simplest one'', does not belong to the pool of gravitational theories available for Nature's selection (except in the Riemannian case).  
In the search for a unique alternative, we have 
considered 
the hypotheses
employed in the formulation of the Lanczos-Lovelock theories of gravity. 
Together, the two approaches point toward
the theory that derives from the Einstein-Hilbert action plus a term quadratic in the torsion vector as the one that would be selected by Nature in the non-Riemannian case without non metricity \textcolor{black}{(when the energy-momentum tensor is identified as its metric version).} The same strategy cannot be followed in the full non-Riemannian case \textcolor{black}{(and in the previous case when the energy-momentum tensor is identified as its canonical version)} as the two approaches are mutually inconsistent.
\end{abstract}
\keywords{Thermodynamics, Non-Riemannian geometry, General relativity, Einstein-Cartan gravity, Lanczos-Lovelock theories of gravity, Poincaré gauge theories of gravity, Metric-affine gravity}
\maketitle

\tableofcontents

\section{Introduction} \label{sec1}

Did Nature have any choice when selecting its own laws?  This is a very profound and interesting question, even more nowadays when a plethora of non-standard scenarios, \textcolor{black}{either well justified or not,} populate the scientific literature.  It is true that well established models and theories exhibit internal inconsistencies, pathologies, or, at best, some lack of capacity to explain key phenomena;\footnote{Some emblematic examples are the following:  the energy to assemble a point charge in classical electromagnetism is infinite \textcolor{black}{(see, anyway, for instance Refs. \cite{Faber:1999ia,Meinert:2024lul})}, the standard model of elementary particles does not explain the origin of neutrino masses \textcolor{black}{(see, anyway, for instance Ref. \cite{Saikumar:2024ahz})}, and the cosmological standard model does not explain the origin and value of the cosmological constant \textcolor{black}{(see, anyway, for instance Ref. \cite{Meinert:2021gpb})}.}  this is, however, a normal characteristic of effective theories.  We should be very careful when proposing new models or theories that, supposedly, supersede the standard ones;  perhaps the latter, despite their shortcomings, are the most favoured by observations in a larger range of applicability (larger than we think) or in a specific framework (e.g., the classical or the quantum frameworks if we are led to consider determinism).  In the specific case of gravity theories, General Relativity (GR) inevitably comes together with singularities \cite{Hawking:1973uf}.  Of course, these singularities do not correspond to physical effects but, instead, signal the limits of the theory.  GR is, then, an effective theory \cite{Burgess:2020tbq,Burgess:2003jk} that, as always has happened, will be superseded by another most powerful one;  but then a key question raises:  is this new theory in the realm of classical physics or does it belong to the quantum framework?  If it belongs to the quantum framework, GR {\em will always be} the best {\em classical} gravity theory at our disposal.  Is this really true?  Did Nature have no choice when selecting the classical gravitational laws? In this paper, we have tried to give a step forward in this direction by assuming two reasonable hypotheses:  the spacetime is described by a smooth manifold $\mathcal{M}$ \cite{lee2003introduction}, which might or might not be Riemannian, and the classical gravitational dynamics is intimately related to the laws of thermodynamics \cite{Jacobson:2003wv,Jacobson:1995ab}.  By analyzing the implications of the thermodynamics' laws, particularly the first one, on the affine and metric structures on $\mathcal{M}$, we now have a panoramic view of the options Nature had.

The relation between gravity and thermodynamics started in the 70's when a set of four laws describing the mechanics of black holes was formulated in \cite{PhysRevD.7.2333,bardeen1973}. These laws were initially derived within the framework of GR and formal analogies to the laws of thermodynamics were established. However, following Hawking's work in \cite{hawking1974,1975CMaPh..43..199H}, it became evident that these laws are, in fact, physically equivalent. This realization prompted Ted Jacobson to pose a fundamental question in \cite{Jacobson:1995ab}: how did GR anticipate that the horizon area would represent entropy and that surface gravity would correspond to temperature?

Jacobson proposed that the area law of entropy, $\delta S \propto \delta \mathcal{A}$ where $S$ is the entropy and $\mathcal{A}$ the area of the horizon, and the Clausius relation, $\delta Q=T \delta S$ where $\delta Q$ is the heat exchange and $T$ is the temperature, are valid for all local Rindler horizons at every point in spacetime \textcolor{black}{(the latter means all processes are quasistatic)}. By identifying $\delta Q$ as the energy flux across the horizon and $T$ as the Unruh temperature perceived by the Rindler observer \cite{PhysRevD.14.870}, Jacobson derived a family of gravitational field equations as thermodynamic \textcolor{black}{fundamental relations}.  This family reduces {\em univocally} to the Einstein field equations of GR after imposing the additional condition of the local conservation of energy-momentum. A crucial assumption in this process is the area law of entropy; this idea is supported by the fact that causal horizons hide information \cite{PhysRevD.7.2333}, implying that they must possess an associated entropy.
Given that the horizon area is an extensive property, it becomes the natural candidate to represent the entropy \textcolor{black}{(what is called the Bekenstein-Hawking entropy)}.

A crucial implication of this thermodynamic derivation, plus the use of the local conservation of the energy-momentum, is that GR may not be a fundamental description of the gravitational interaction. Instead, it can be interpreted as \textcolor{black}{a fundamental relation} for a thermodynamic system of space-time microstates 
under conditions of local equilibrium. This insight led Eling, Guedens, and Jacobson \cite{Eling:2006aw} to successfully extend this framework to $f(R)$ gravity, this time \textcolor{black}{generalizing the Clausius relation to allow for non-equilibrium terms}.  These results suggest that the thermodynamic description might be an intrinsic property of gravitation. 

Jacobson's results were obtained by assuming that spacetime is a Riemannian manifold \cite{lee2018introduction}, i.e., it is a manifold whose affine structure is characterized by a connection that is symmetric and compatible with the metric structure.  This is the usual assumption when constructing gravitational theories and it is the one employed, indeed, when constructing GR (in part motivated by the equivalence principle) \cite{thorne2000gravitation}.  What happens then when this assumption is relaxed by making the connection asymmetric, incompatible with the metric structure, or both?  The asymmetry of the connection is characterized by a tensor called Torsion whereas the incompatibility of the connection with the metric structure is characterized by a tensor called Non Metricity.  The extension of Jacobson's results to a spacetime with an asymmetric but metric-compatible connection was explored in \cite{Dey:2017fld,DeLorenzo:2018odq} and it was claimed that the Einstein-Cartan (EC) theory \cite{ASENS_1923_3_40__325_0} is one of the possible realizations. This seems to further support Jacobson's approach;  indeed, the EC theory, whose action is the Einstein-Hilbert one, is ``the simplest'' model within the set of gravitational theories that involve both curvature and torsion but whose connection is metric compatible;  this set is called the Poincaré Gauge Theories of Gravity (PGT) \cite{blagojevic2001gravitation,Blagojevic:2013xpa,Hehl:1976kj,Hehl1979}. In the PGT framework, the gravitational models are constructed by localizing the Poincaré group of transformations, in complete analogy with the construction of the other fundamental interactions. Within PGT, torsion acts as a necessary field associated with the translation subgroup, sourced directly by the spin of matter;  consequently, GR is recovered as the specific limit of the EC theory where the particles' microstructure is ignored and, therefore, the torsion tensor vanishes.

These findings suggest that gravity is an emergent phenomenon resulting from the dynamics of spacetime's fundamental structure \cite{Smolin:2012ys,Chirco:2014saa}. Researchers have been thus applying statistical mechanics to discrete space-time models to recover GR \cite{Gielen:2013kla,Oriti:2015qva,Oriti:2015rwa,Oriti:2016qtz,Dittrich:2013xwa,Dittrich:2014mxa,Dittrich:2016tys}. As this emergent perspective allows for theories beyond GR \cite{Eling:2006aw,Dey:2016zka,Guedens:2011dy}, Jacobson's procedure may prove essential for identifying the best classical theory of gravity.

On the other hand, the Poincaré group can be extended to include general linear transformations; this corresponds to the affine group, and its localization, known as Metric-Affine Gravity (MAG) \cite{Blagojevic:2013xpa,Hehl:1994ue,JimenezCano:2021rlu}, provides a new framework for studying gravitational phenomena. In MAG, non metricity acquires a physical role alongside the torsion field, requiring the use of the most general linear connection. Consequently, the energy-momentum tensor is accompanied by \textcolor{black}{spin, dilatation, and shear currents} \cite{Andrei:2024vvy}. These matter field currents act as new sources for the gravitational field and are associated with the dynamics of the general connection.

It is worthwhile mentioning that the study of these scenarios is highly relevant for solid-state physics. While deformations of ideal crystals can be described using the standard techniques of GR (metric curvature), real crystals contain local imperfections such as point defects and dislocations. Describing these features requires geometric structures beyond the metric \textcolor{black}{tensor} \cite{falk1981,Kupferman2015,Kupferman2017}. By analogy, this suggests that the affine connection encodes information about microscopic defects in the space-time structure, defects that the metric tensor alone cannot describe.

In this work, we have investigated the general case of a non-Riemannian spacetime (involving curvature, torsion, and non metricity), \textcolor{black}{assuming local thermodynamic equilibrium.} 
Our results demonstrate that the Einstein-Hilbert action {\em does not remain as one of the alternatives} Nature had when selecting the classical gravitational theory (except in the Riemannian case).  Of course, ``more complex'' possibilities exist. 
Following the Ockham's razor philosophy, we have employed the hypotheses in the construction of the Lanczos-Lovelock theories \cite{Lanczos:1932fxw,Lanczos:1938sf,Lovelock:1969vyr,Lovelock:1970zsf,Lovelock:1971yv,Lovelock:1972vz} to 
reduce the spectrum of possibilities in the Jacobson's framework to just one.\footnote{The local conservation of the energy-momentum \textcolor{black}{tensor} that Jacobson employed in his pioneering work \cite{Jacobson:1995ab} to univocally obtain GR is, indeed, one of the Lanczos-Lovelock hypotheses.}  As the first main conclusion of our paper, we have found out the {\em unique} gravitational theory that meets the Jacobson and Lanczos-Lovelock's requirements for a non-Riemannian spacetime without non metricity \textcolor{black}{when the energy-momentum tensor is identified as its metric version}: {\em it is the one that derives from the Einstein-Hilbert action plus a term quadratic in the torsion vector}. The second main conclusion has arisen {\em when non metricity has shown up \textcolor{black}{or in the previous case when the energy-momentum tensor is identified as its canonical version}: these sets of requirements mutually conflict}.  As a byproduct of our investigation, we have identified the new terms associated with the non metricity that could contribute to the Hartle-Hawking tidal heating of black holes; this finding might give hints toward the search for observational evidence of torsion and non metricity.

This document is organized as follows: in Section \ref{sec3}, we briefly review some key aspects of the non-Riemannian geometry. In Section \ref{sec2}, we review the rationale behind the derivation of GR in the framework of the Jacobson's ideas and apply it to a non-Riemannian spacetime whose connection is asymmetric but compatible with the metric structure. In Section \ref{sec4}, we derive the most general gravitational fundamental relation following Jacobson's approach in a general non-Riemannian spacetime and apply on it the Lanczos-Lovelock hypotheses. 
Finally, in Section \ref{sec7}, we discuss the implications of these results. Definitions, a proof, and some detailed derivations are presented in the appendices.  
Throughout the paper, we have assumed the spacetime is four dimensional\footnote{One of the authors (Yeinzon Rodríguez) has the strong belief that there are no enough reasons, either theoretical or empirical, to contemplate the possibility of extra spatial dimensions \textcolor{black}{(see, for instance, Ref. \cite{Pardo:2018ipy})}.} and employed the mostly positive signature.

\section{Non-Riemannian Geometry} \label{sec3}
Recall that we aim to study the most general gravitational fundamental relation in a non-Riemannian spacetime $\mathcal{M}$.  By definition, this includes the (Lorentzian) metric $g$, torsion $T$, and non metricity $Q$. We work within thermodynamic equilibrium \textcolor{black}{so that} the Clausius relation holds. With this objective \textcolor{black}{in mind}, our first task is to review the basics of metric-affine geometry by outlining its key features.

A general affine connection can be described by its coefficients denoted as $\Gamma^{\alpha}{}_{\mu\nu}$. 
These can be expressed as \cite{schouten2013ricci}
\begin{equation}
\Gamma^{\alpha}_{\hspace{2mm}\mu \nu}=\mathring{\Gamma}^{\alpha}{}_{\mu \nu}+\tilde{\Gamma}^{\alpha}{}_{\mu \nu} \,,
\label{MAG01}
\end{equation}
where, in a coordinate base, 
\begin{equation}
\mathring{\Gamma}^{\nu}_{\hspace{2mm}\alpha \mu}=\frac{1}{2}g^{\nu \beta}(\partial_{\alpha}g_{\beta \mu}+\partial_{\mu}g_{\alpha \beta}-\partial_{\beta}g_{\mu \alpha}) \,,
\label{MAG02}
\end{equation}
are the connection coefficients induced by the metric \textcolor{black}{tensor}, i.e., the components of the Levi-Civita connection, and
\begin{align}
\tilde{\Gamma}^{\nu}_{\hspace{2mm}\alpha \mu}=& \frac{1}{2}(T^{\nu}{}_{\alpha \mu}-T_{\alpha \mu}{}{}^{\nu}-T_{\mu \alpha}{}{}^{\nu}) \nonumber \\
& +\frac{1}{2}(Q^{\nu}{}_{\alpha \mu}-Q_{\mu \alpha}{}{}^{\nu}-Q_{\alpha \mu}{}{}^{\nu}) \,,
\label{MAG03}
\end{align}
are the components of the distortion tensor.  These, in turn, are given in terms of the components of the torsion tensor
\begin{equation}
T^{\nu}{}_{\alpha \mu}\equiv{\Gamma}^{\nu}_{\hspace{2mm}\alpha \mu}-{\Gamma}^{\nu}{}_{\mu \alpha}=\tilde{\Gamma}^{\nu}{}_{\alpha \mu}-\tilde{\Gamma}^{\nu}{}_{\mu \alpha} \,,
\label{MAG04}
\end{equation}
and the components of the non-metricity tensor
\begin{equation}
Q_{\mu \alpha \beta}\equiv\nabla_{\mu}g_{\alpha \beta} \,,
\label{MAG05}
\end{equation}
where, in the previous expression, $\nabla_\mu$ represents the components of the covariant derivative associated to the general affine connection.
We denote quantities with an over-circle, e.g., $\mathring{\Gamma}, \mathring{\nabla}$, as objects defined with respect to the Levi-Civita connection.
Therefore, a metric-affine spacetime is a set ($\mathcal{M}, g, T, Q, \epsilon$), where $\epsilon$ is the metric volume form, together with an arrow of time. Appendix \ref{appendix1} contains additional definitions and properties in a non-Riemannian spacetime.

We now introduce some vector quantities that will be useful for subsequent calculations. First, we can construct a vector from the torsion tensor whose components are
\begin{equation}
T_{\mu}\equiv T^{\nu}{}_{\nu \mu} \,.
\label{MAG08}
\end{equation}
We also define two vectors associated with the non metricity whose components are
\begin{equation}
Q_{\mu}\equiv Q^{\nu}{}_{\nu \mu} \,,
\label{MAG09}
\end{equation}
and
\begin{equation}
\tilde{Q}_{\mu}\equiv Q_{\mu\nu}{}{}^{\nu} \,.
\label{MAG10}
\end{equation}

On the other hand, a Killing vector field $\xi$ is one along which the metric tensor is preserved. It generates isometries within the given manifold meaning that the Lie derivative of the metric tensor along the Killing vector field is zero:
\begin{equation}
\mathcal{L}_{\xi}g_{\mu \nu}=0 \,.
\label{MAG11}
\end{equation}
This can be written in terms of partial derivatives of the metric tensor as
\begin{equation}
\xi^{\rho}\partial_{\rho}g_{\mu \nu}+g_{\rho \nu} \partial_{\mu} \xi^{\rho}+g_{\rho \mu}\partial_{\nu}\xi^{\rho}=0 \,,
\label{MAG12}
\end{equation}
or in terms of the affine connection as
\begin{equation}
g_{\rho \nu} \nabla_{\mu} \xi^{\rho}+g_{\rho \mu}\nabla_{\nu}\xi^{\rho}+
\xi^{\rho}Q_{\rho\mu \nu}+\xi^{\rho}\Big(T_{\nu \rho \mu}+T_{\mu \rho \nu} \Big)=0 \,.
\label{MAG13}
\end{equation}
Other useful relations for $\xi$, coming from its definition, are the following:
\begin{align}
& 2\nabla_{\nu}\xi^{\nu}+\xi^{\mu}\tilde{Q}_{\mu}-2\xi^{\mu}T_{\mu}=0 \,,
\label{eqA14} \\
& \xi^{\mu}\nabla_{\mu}\xi_{\nu} +\xi_{\mu}\nabla_{\nu}\xi^{\mu}+\xi^{\mu}\xi^{\rho}Q_{\rho \mu \nu}+\xi^{\mu}\xi^{\rho}T_{\mu \rho \nu}=0 \,,
\label{eqA15} \\
& 2\xi_{\mu}\xi^{\nu}\nabla_{\nu}\xi^{\mu}+\xi^{\mu}\xi^{\nu}\xi^{\rho}Q_{\rho \mu \nu}=0 \,.
\label{eqA16}
\end{align}

\subsection{Raychaudhuri equation for a null congruence}
Let us consider the Rindler horizon we are interested in.  It is a null hypersurface $\mathscr{H}$, i.e., a three-dimensional submanifold with a future-directed null normal vector field denoted by $l$. The hypersurface can be defined as a level set of a scalar field $\varphi(x^\mu)=0$. The normal vector is collinear \textcolor{black}{to the dual of the gradient of the scalar field $\varphi$ by metric duality:}
\begin{equation}
 l^\mu=-\exp{(\rho)} \ g^{\mu\nu}\nabla_\nu \varphi \,,   
\end{equation}
where $\rho$ is some scalar field and the sign is chosen so that $l$ is future directed. 
The hypersurface $\mathscr{H}$ is generated by a null congruence of curves with tangent vector $l$ and parameter $\lambda$. It is worthwhile determining whether this congruence is geodesic;  to this end, we evaluate the commutator
\begin{equation}
l^\alpha\nabla_\alpha l_\beta -l^{\alpha} \nabla_\beta l_\alpha =l^\alpha  l_\beta \nabla_\alpha \rho+l^\alpha  l^\gamma  T_{\alpha \beta \gamma} \,. \label{eqn:commutator_l}   
\end{equation}
Since $l$ is null, 
\begin{equation}
    \nabla_\beta(l_\alpha l^\alpha) = 2l^\alpha\nabla_\beta l_\alpha - l^\alpha l^\gamma Q_{\beta\alpha\gamma} = 0 \,;
\end{equation}
thus, Eq. \eqref{eqn:commutator_l} becomes
\begin{equation}
    l^\alpha\nabla_\alpha l_\beta = l^\alpha  l_\beta \nabla_\alpha \rho+l^\alpha  l^\gamma  T_{\alpha \beta \gamma } + \tfrac{1}{2}l^\alpha l^\gamma Q_{\beta\alpha\gamma} \,.
\end{equation}
This demonstrates that, in the absence of torsion and non metricity, the congruence reduces to a geodesic flow.

We will now study the general Raychaudhuri equation in the presence of both torsion and non metricity. There exists a two-dimensional submanifold $\mathscr{L}$ called the cross-section of $\mathscr{H}$ which is a space-like hypersurface of $\mathscr{H}$. The normal vector $l$ is not tangent to $\mathscr{L}$ and each generating null curve intersects $\mathscr{L}$ at most once.  There exists a unique future-directed null vector $k$ at each point in $\mathscr{L}$, transverse to $\mathscr{H}$ such that $k_\mu l^\mu = -1$.  Thus, the metric \textcolor{black}{tensor} induced  in the cross-section $\mathscr{L}$ is given by
\begin{equation}
q_{\mu \nu} = g_{\mu \nu}+l_{\mu}k_{\nu}+l_{\nu}k_{\mu} \,.
\label{RAY02} 
\end{equation}

The expansion $\Theta$ of the cross-section $\mathscr{L}$ along the vector field $l$ is given by
\begin{equation}
\Theta = \frac{1}{2}q^{\mu \nu}\mathcal{L}_{l}q_{\mu \nu} \,.
\label{RAY03}
\end{equation}
Employing Eq. \eqref{MAG13}, the previous expression can be written as\footnote{In Ref. \cite{Dey:2017fld}, the authors used an affinely parameterized null horizon generator, implying an extra condition on the torsion tensor. This was done because the authors wanted to obtain a generalized version of the thermodynamics' zeroth law. In this paper, we have not employed such a condition and this is the reason why Eq. (\ref{RAY04}), in the limit when $Q = 0$, is not the same as Eq. (57) of Ref. \cite{Dey:2017fld}.}
\begin{align}
\Theta =  &\nabla_{\mu}l^{\mu}-l^{\mu}T_{\mu}+\frac{1}{2}l^{\mu}\tilde{Q}_{\mu}+k^{\mu}l_{\nu}\nabla_{\mu}l^{\nu} \nonumber \\
&+k^{\mu}k^{\nu}l^{\rho}Q_{\rho \mu \nu}+k^{\mu}l^{\nu}l^{\rho}T_{\nu \rho \mu}+k_{\mu} l^{\nu}\nabla_{\nu} l^{\mu} \,.
\label{RAY04}
\end{align}

The evolution of the expansion along the vector $l$ is defined as $d\Theta/d\lambda = l^\mu\nabla_\mu\Theta$. We define $\mathcal{P}$ as a point in $\mathscr{L}$ but otherwise arbitrary.  Since $l$ is a Killing vector for the submanifold $\mathscr{L}$, the expansion vanishes at $\mathcal{P}$: $\Theta|_{\mathcal{P}}=0$. Therefore,
\begin{align}
\frac{d\Theta}{d\lambda}\Big|_{\mathcal{P}} = & l^{\alpha}\nabla_{\alpha}\nabla_{\mu}l^{\mu}+ l^{\nu}\nabla_{\nu} l^{\mu}\Big(\frac{1}{2}\tilde{Q}_{\mu}-T_{\mu}\Big) \nonumber \\
& +l^{\mu}l^{\nu}\Big(\frac{1}{2}\nabla_{\nu}\tilde{Q}_{\mu}-\nabla_{\nu}T_{\mu}\Big) \,.
\label{RAY06}
\end{align}
Using Eq. (\ref{eqA04}) and a total derivative, we can write
\begin{align}
\frac{d\Theta}{d\lambda}\Big|_{\mathcal{P}} =& -\nabla_{\mu}\Big(l_{\rho}\nabla^{\mu}l^{\rho}+l_{\nu}l_{\rho}Q^{\rho\nu\mu}+l_{\nu}l_{\rho}T^{\nu\rho\mu}\Big) \nonumber \\
& -\Big(l_{\rho}\nabla^{\mu}l^{\rho}+l_{\nu}l_{\rho}Q^{\rho\nu\mu}+l_{\nu}l_{\rho}T^{\nu\rho\mu}\Big)\Big(\frac{1}{2}\tilde{Q}_{\mu}-T_{\mu}\Big) \nonumber \\
& -\nabla_{\mu}l^{\alpha}\nabla_{\alpha}l^{\mu}-l^{\mu}l^{\nu}R_{\mu \nu} \nonumber \\
&+l^{\mu}l^{\nu}\Big(\frac{1}{2}\nabla_{\nu}\tilde{Q}_{\mu}-\nabla_{\nu}T_{\mu}\Big)+T_{\beta \mu \alpha}l^{\alpha}\nabla^{\beta}l^{\mu} \,,
\label{RAY08}
\end{align}
where $R_{\mu \nu}$ are the components of the Ricci tensor.
Thus, we can write\footnote{Ref. \cite{Iosifidis:2018diy} presented a generalized version of Raychaudhuri's equation for a general non-Riemannian $n$-dimensional spacetime. The expansion in that work was defined as $\Theta \equiv \nabla^{\nu} u_{\nu}$, where $u_{\nu}$ represents the components of the $n$-velocity vector tangent to the respective congruence of time-like curves. In our case, we want to describe the expansion experienced by a two-dimensional space-like hypersurface through which heat is flowing and this is the reason why Eq. (\ref{RAY10}) is not the same as Eq. (41) of Ref. \cite{Iosifidis:2018diy}.}
\begin{widetext}
\begin{align}
\frac{d\Theta}{d\lambda}\Big|_{\mathcal{P}} =& -R_{\alpha \beta}l^{\alpha}l^{\beta}-l^{\alpha}l^{\beta}\nabla_{\alpha}T_{\beta}+l^{\alpha}l^{\beta}T_{\alpha \beta \nu}T^{\nu}-l^{\alpha}l^{\beta}\nabla_{\nu}T_{\beta \alpha}^{\hspace{4mm}\nu}
+\frac{1}{2}l^{\alpha}l^{\beta}\nabla_{\alpha}\tilde{Q}_{\beta}-\frac{1}{2}l^{\alpha}l^{\beta}Q_{\alpha \beta \nu}\tilde{Q}^{\nu} \nonumber \\
&-l^{\alpha}l^{\beta}\nabla_{\nu}Q_{\beta \alpha}{}{}^{\nu}
+l^{\alpha}l^{\beta}T^{\nu}Q_{\alpha \beta \nu}-\frac{1}{2}l^{\alpha}l^{\beta}T_{\alpha \beta \nu}\tilde{Q}^{\nu}
-(\nabla_{\beta}l^{\alpha})(\nabla_{\alpha}l^{\beta})
-T_{\mu \rho \nu}\nabla^{\nu}(l^{\mu}l^{\rho}) \nonumber \\
&+T_{\beta \mu \alpha}l^{\alpha}\nabla^{\beta}l^{\mu}+T^{\nu}l_{\mu}\nabla_{\nu}l^{\mu}-\frac{1}{2}\tilde{Q}_{\nu}l_{\mu}\nabla^{\nu}l^{\mu}-Q_{\rho \mu \nu}\nabla^{\nu}(l^{\mu}l^{\rho})-\nabla_{\nu}(l_{\mu}\nabla^{\nu}l^{\mu}) \,.
\label{RAY10}
\end{align}
\end{widetext}
\section{Jacobson's Thermodynamic Approach:  The Simplest Cases} \label{sec2}
We will now briefly review Jacobson's thermodynamic approach applied to a Riemannian spacetime as well as to a non-Riemannian spacetime whose connection is asymmetric but compatible with the metric structure.  As a result, we will obtain a selected family of gravity theories in each case.  This family  includes GR in the first case \cite{Jacobson:1995ab} but, contrary to what was claimed in \cite{Dey:2017fld,DeLorenzo:2018odq}, does not include the EC theory in the second case.  After introducing the Lanczos-Lovelock hypotheses, we will see that, whereas the gravitational theory that derives from the Einstein-Hilbert action {\em is} selected as the unique gravitational theory in the first case, \textcolor{black}{a slight deviation from this action gives way to the selected gravitational theory in the second case.  This deviation is quadratic on the torsion vector and is valid only when the energy-momentum tensor is identified as its metric version;  in contrast, if the energy-momentum tensor is identified as its canonical version, there is no way to derive the field equation obtained from the Jacobson's thermodynamic approach from an action principle.}

\subsection{Riemannian Spacetime} \label{riem}
In a Riemannian spacetime, we consider a local Rindler horizon generated by an approximate local boost Killing vector field $\chi$. {\em The heat flux is identified with the flux of boost energy across the horizon}:
\begin{equation}
\delta Q=\int_{\mathscr{H}}\tau_{\mu \nu}\chi^{\mu}d\Sigma^{\nu} \,,
\label{GR01}    
\end{equation}
where $\tau_{\mu \nu}$ are the components of the metric energy-momentum tensor associated with the matter fields:
\begin{equation}
\tau_{\mu \nu}=-\frac{2}{\sqrt{-\det g}}\frac{\delta(\sqrt{-\det g} \mathcal{L}_{M})}{\delta g^{\mu \nu}} \,.
\label{GR08}
\end{equation}
Here, the matter Lagrangian density $\mathcal{L}_M$ is a function of just the metric, the Levi-Civita connection, and the matter fields,
whereas $d\Sigma^\nu$ are the components of a vector field orthogonal to $\mathscr{H}$, proportional to the volume of the respective differential section of $\mathscr{H}$, and chosen so that $\epsilon(d\Sigma,\cdot,\cdot,\cdot)$ is compatible with the orientation of $\mathscr{H}$. Since $l$ is the vector field that is tangent to the horizon generators corresponding to an affine parameter $\lambda$, we can rewrite $\chi^{\alpha}=-\kappa \lambda l^{\alpha}$, where $\kappa$ is the norm of the observer's four-acceleration, i.e., it is a constant, and $d\Sigma^{\nu}=l^{\nu}d\lambda d\mathcal{A}$ so that
\begin{equation}
\delta Q=-\kappa\int_{\mathscr{H}}\lambda \tau_{\mu \nu}l^{\mu}l^{\nu}d\lambda d\mathcal{A} \,.
\label{GR02}
\end{equation}

The variation of the horizon area is given by
\begin{equation}
\delta\mathcal{A}=\int_{\mathscr{H}}\Theta d\lambda d\mathcal{A} \,,
\label{GR03}
\end{equation}
where $\Theta$ is the expansion of the null congruence. Thus, from the definition of expansion, we have
\begin{align}
\begin{aligned}
\frac{d\Theta}{d\lambda}&=l^{\mu}\mathring{\nabla}_{\mu}\Theta=l^{\mu}\mathring{\nabla}_{\mu}(\mathring{\nabla}_{\nu} l^{\nu}) \\
&=-\mathring{R}_{\alpha \beta}l^{\alpha}l^{\beta}+l^{\alpha}\mathring{\nabla}_{\mu}\mathring{\nabla}_{\alpha}l^{\mu} \,.
\end{aligned}
\label{GR04}
\end{align}
Upon Taylor expanding $\Theta$ in Eq. (\ref{GR03}) and neglecting contributions of order $\mathcal{O}(\lambda^2)$ and higher, we obtain $\Theta \approx -\lambda \mathring{R}_{\alpha \beta}l^{\alpha}l^{\beta}$. \footnote{The contribution to the change in the entropy  
\begin{equation}
\delta S_i=-\eta \int_{\mathscr{H}}(\mathring{\nabla}_{\mu}l^{\alpha})(\mathring{\nabla}_{\alpha}l^{\mu})\lambda d\lambda d\mathcal{A} \,, \nonumber
\end{equation}
 contains quadratic terms in the shear and vorticity. It was identified as a non-equilibrium contribution in Jacobson's subsequent works \cite{Eling:2006aw,Guedens:2011dy}. \label{neqs}}

Now, according to the area law of entropy \textcolor{black}{(Bekenstein-Hawking entropy)}, $\delta S= \eta \delta \mathcal{A}$, where $\eta$ is a constant. Invoking the fundamental equilibrium relation $\delta Q = T \delta S$, we find that $\delta Q= \eta\, T\, \delta \mathcal{A}$.  On the other hand, the Unruh effect states that the vacuum state of quantum fields, as measured by the uniformly accelerated observer, has a temperature $T=\hbar \kappa/2\pi$, where $\hbar$ is the reduced Planck constant \cite{PhysRevD.14.870}. Substituting the previous results yields
\begin{equation}
\kappa\int_{\mathscr{H}}\lambda \tau_{\mu \nu}l^{\mu}l^{\nu}d\lambda d\mathcal{A}=\frac{\eta \hbar \kappa}{2 \pi}\int_{\mathscr{H}}\lambda \mathring{R}_{\alpha \beta}l^{\alpha}l^{\beta} d\lambda d\mathcal{A} \,.
\label{GR05}
\end{equation}

Since this relation must be valid for all local Rindler horizons, it implies 
\begin{equation}
\tau_{\alpha \beta}l^{\alpha}l^{\beta}=(\eta \hbar /2 \pi) \mathring{R}_{\alpha \beta}l^{\alpha}l^{\beta} \,. \label{tommy}
\end{equation}
Consequently,
\begin{equation}
\boxed{\frac{2\pi}{\hbar \eta}\tau_{\alpha \beta}=\mathring{R}_{\alpha \beta}+f g_{\alpha \beta} \,,}
\label{GR06}
\end{equation}
where $f$ is an unknown scalar field.\footnote{The second term in the right member of Eq. (\ref{GR06}) should be written, in principle, as a symmetric tensor $h$ that satisfies $h_{\alpha \beta} l^\alpha l^\beta = 0$ for any null vector $l$.  It turns out that $h_{\alpha \beta} = f g_{\alpha \beta}$ as demonstrated in Appendix \ref{fg}.}  Eq. (\ref{GR06}) identifies the family of gravity theories that are compatible with Jacobson's thermodynamic approach in a Riemannian spacetime and is, in consequence, the most important equation in this subsection.  The physical interpretation of Jacobson's approach is that the space-time curvature reacts to the presence of matter in a particular way, that described by Eq. (\ref{GR06}), in order not to violate the laws of thermodynamics \cite{Jacobson:1995ab}.

The family described by Eq. (\ref{GR06}) contains an infinite number of members.  Jacobson employed the contracted Bianchi identity and assumed diffeomorphism invariance of the action (which in the context of a Riemannian geometry implies local conservation of the energy-momentum \textcolor{black}{tensor}: $\mathring{\nabla}_{\mu}\tau^{\mu \nu}=0$);  thus, he found $f = -\mathring{R}/2+\Lambda$ where $R$ is the Ricci scalar and $\Lambda$ is a constant. It then follows that
\begin{empheq}[box=\fbox]{equation}
\mathring{R}_{\mu \nu}-\frac{1}{2}g_{\mu \nu}\mathring{R}+\Lambda g_{\mu \nu}=8\pi G \tau_{\mu \nu} \,,
\label{GR07}
\end{empheq}
which coincides with the Einstein field equations. We have set $\eta=(4\hbar G)^{-1}$, where $G$ is the universal gravitation constant, to recover Newtonian gravity in the weak-field limit; this implies $\eta^{-1/2}$ corresponds to twice the Planck length. Notably, the cosmological constant $\Lambda$ appears as an integration constant, though its nature remains unknown.  
It is worthwhile remarking that the {\em uniqueness} of Eq.~\eqref{GR07} (except for the, a priori, arbitrary value of $\Lambda$) is guaranteed by the local conservation of energy-momentum, i.e., by one of the Lanczos-Lovelock's hypotheses \cite{Lanczos:1932fxw,Lanczos:1938sf,Lovelock:1969vyr,Lovelock:1970zsf,Lovelock:1971yv,Lovelock:1972vz}.  The other three hypotheses:  that the right member of Eq. (\ref{GR06}) i) is built out from the metric tensor and its first- and second-order derivatives, ii) is symmetric, and iii) is linear in second-order derivatives of the metric \textcolor{black}{tensor}, are easily seen to be satisfied.

\textcolor{black}{Before finishing this subsection, let us reflect a bit about the procedure we have followed.} While black hole thermodynamics motivates the holographic hypothesis \cite{tHooft:1993dmi,Susskind:1994vu} and supports the entropy-area proportionality, relying on it would be circular since those results typically presume the validity of GR. Instead, \textcolor{black}{we have followed} the insight in \cite{PhysRevD.7.2333}: causal horizons hide information which implies they must possess an associated entropy. This hidden information resides in the correlations between vacuum fluctuations just inside and outside the horizon \cite{PhysRevD.34.373}.

\subsection{Non-Riemannian Spacetime with Vanishing Non Metricity} \label{nriemvnm}
We will now work in a non-Riemannian spacetime whose connection is asymmetric but compatible with the metric structure.  The procedure is the same as in the Riemannian case except for the facts that the diffeomorphism invariance of the action no longer guarantees the conservation of the metric energy-momentum, that the expansion of the null congruence is modified, \textcolor{black}{and that we have to consider the difference between the canonical and metric energy-momentum tensors which, for this setup, are not identical.}  The origin of these three exceptions lies in the existence of torsion.

When the non metricity vanishes, the Raychaudhuri equation in \eqref{RAY10} reduces to 
\begin{align}
\frac{d\Theta}{d\lambda}\Big|_{\mathcal{P}} = &-R_{\alpha \beta}l^{\alpha}l^{\beta}-l^{\alpha}l^{\beta}\nabla_{\alpha}T_{\beta}+l^{\alpha}l^{\beta}T_{\alpha \beta \nu}T^{\nu} \nonumber \\
&-l^{\alpha}l^{\beta}\nabla_{\nu}T_{\beta \alpha}{}{}^{\nu} -T_{\mu \rho \nu}\nabla^{\nu}(l^{\mu}l^{\rho}) \nonumber \\
&-(\nabla_{\mu}l^{\alpha})(\nabla_{\alpha}l^{\mu})+T_{\beta \mu \alpha}l^{\alpha}\nabla^{\beta}l^{\mu} \,.\label{EC01}
\end{align}
Upon Taylor expanding $\Theta$ in Eq. (\ref{GR03}) and neglecting contributions of order $\mathcal{O}(\lambda^2)$ and higher, we end up with\footnote{The non-equilibrium contribution to $\delta S$ is, in this case,
\begin{align}
\delta S_i=& \eta \int_{\mathscr{H}}\Big(-T_{\mu \rho \nu}\nabla^{\nu}(l^{\mu}l^{\rho})-(\nabla_{\mu}l^{\alpha})(\nabla_{\alpha}l^{\mu}) \nonumber \\
&+T_{\beta \mu \alpha}l^{\alpha}\nabla^{\beta}l^{\mu} \Big)\lambda d\lambda d\mathcal{A} \nonumber \,,
\end{align}
as noted in \cite{Dey:2017fld} (except for an ambiguous term which was added up there without enough justification). This contribution contains quadratic terms in the shear and vorticity.  We can check that, when torsion vanishes, this contribution reduces to the one presented in footnote \ref{neqs}. \label{foot6}}
\begin{equation}
\Theta \approx \lambda (-R_{\alpha \beta}l^{\alpha}l^{\beta}-l^{\alpha}l^{\beta}\nabla_{\alpha}T_{\beta}+l^{\alpha}l^{\beta}T_{\alpha \beta \nu}T^{\nu}
 -l^{\alpha}l^{\beta}\nabla_{\nu}T_{\beta \alpha}^{\hspace{3mm}\nu}) \,.
\end{equation}

Thus, by invoking the area law of entropy, $\delta S = \eta \delta \mathcal{A}$, and the Unruh temperature, $T = \hbar \kappa /2\pi$, the Clausius relation $\delta Q = T \delta S$ acquires the form
\begin{align}
\kappa\int_{\mathscr{H}}\lambda \textcolor{black}{\mathcal{T}_{\mu \nu}}l^{\mu}l^{\nu}d\lambda d\mathcal{A}= & \frac{\eta \hbar \kappa}{2 \pi}\int_{\mathscr{H}} \lambda (R_{\alpha \beta}l^{\alpha}l^{\beta} +l^{\alpha}l^{\beta}\nabla_{\alpha}T_{\beta} \nonumber \\
&-l^{\alpha}l^{\beta}T_{\alpha \beta \nu}T^{\nu} \nonumber \\
&+l^{\alpha}l^{\beta}\nabla_{\nu}T_{\beta \alpha}^{\hspace{3mm}\nu}) \ d\lambda d\mathcal{A} \,, \label{tauorsigma}
\end{align}
\textcolor{black}{where $\mathcal{T}$ is the energy-momentum tensor.}  

This relation being valid for all local Rindler horizons, we have to conclude that
\begin{align}
\frac{2\pi}{\hbar \eta}\textcolor{black}{\mathcal{T}_{\alpha \beta}}l^{\alpha}l^{\beta}=& R_{\alpha \beta}l^{\alpha}l^{\beta}+l^{\alpha}l^{\beta}\nabla_{\alpha}T_{\beta}-l^{\alpha}l^{\beta}T_{\alpha \beta \nu}T^{\nu} \nonumber \\
& +l^{\alpha}l^{\beta}\nabla_{\nu}T_{\beta \alpha}^{\hspace{3mm}\nu} \,,
\label{EC02}
\end{align}
and, therefore,
\begin{equation}
\frac{2\pi}{\hbar \eta}\textcolor{black}{\mathcal{T}_{(\alpha \beta)}} = R_{(\alpha \beta)}+\nabla_{(\alpha}T_{\beta)}-T_{(\alpha \beta) \nu}T^{\nu}+\nabla_{\nu}T_{(\beta \alpha)}^{\hspace{5mm}\nu} +f g_{\alpha \beta} \,,
\end{equation}
where $f$ is an unknown scalar field and the brackets that enclose indices mean normalized symmetrization.  
This expression may be written more elegantly as 
\begin{empheq}[box=\fbox]{equation}
\frac{2\pi}{\hbar \eta}\textcolor{black}{\mathcal{T}_{(\alpha \beta)}} = R_{(\alpha \beta)} +f g_{\alpha \beta} +(\nabla_{\nu}-T_{\nu})T_{(\alpha \beta)}{}{}{}{}^{\nu} +\nabla_{(\alpha} T_{\beta)} \,.
\label{EC05}
\end{empheq}
Eq. (\ref{EC05}) identifies the family of gravity theories that are compatible with Jacobson's thermodynamic approach in a non-Riemannian spacetime with vanishing non metricity and is, in consequence, one of the most important equations in this subsection.

\textcolor{black}{There exists, however, an ambiguity regarding the identity of $\mathcal{T}$. In the Riemannian case, the metric energy-momentum tensor $\tau$ and the canonical one $\Sigma$ are just the same when the action is invariant under both diffeomorphisms and local Poincaré gauge transformations;\footnote{\textcolor{black}{See an interesting discussion on the difference between $\tau$ and $\Sigma$ in the Riemannian case in \cite{Blaschke:2016ohs,Baker:2020eqs}.}}  this is no longer the case, however, when the spin density is different to zero.  Which energy-momentum tensor should we identify then as $\mathcal{T}$ in Eq. (\ref{tauorsigma}): $\tau$ or $\Sigma$?}

\subsubsection{\textcolor{black}{$\mathcal{T} \equiv \tau$}}

\textcolor{black}{Let us suppose that $\mathcal{T}$ is identified as the metric energy-momentum tensor $\tau$.}
It is well known that both GR and the EC theory derive from the Einstein-Hilbert action.  However, whereas the action is formulated on a Riemannian spacetime for the GR case, it is formulated on a non-Riemannian spacetime without non metricity for the EC case.  In the EC theory, there exist two gravitational field equations \cite{ASENS_1923_3_40__325_0,blagojevic2001gravitation,Blagojevic:2013xpa,Hehl:1976kj,Hehl1979};  \textcolor{black}{the symmetric part of} one of them is
\begin{align}
R_{(\mu \nu)}-\frac{1}{2}g_{\mu \nu}(R-2\Lambda)=& 8\pi G \tau_{\mu \nu} -(\nabla_{\sigma}-T_{\sigma}) \Big(T_{(\mu \nu)}{}{}{}{}^{\sigma} \nonumber \\
&+ \delta^\sigma_{(\mu} T_{\nu)} -g_{\mu \nu}T^{\sigma} \Big) \,,
\label{EC06}
\end{align}
(see Appendix \ref{appendix2} in the limit when $Q = 0$) and the other is just a constraint equation involving the torsion and the hypermomentum tensor (which, in this case, just contains information about the spin \cite{Andrei:2024vvy}). By comparing Eq. (\ref{EC06}) with Eq. (\ref{EC05}), \textcolor{black}{with $\mathcal{T} \equiv \tau$,} it is evident that the EC theory does not belong to the family of gravitational theories selected by Jacobson's thermodynamic approach.

We have to keep in mind that, in contrast to the Riemannian case,  \textcolor{black}{the invariance of the action under diffeomorphisms does not lead to the local conservation of the energy-momentum tensor.}  Then, what is the generalization of the respective Lanczos-Lovelock hypothesis for non-Riemannian spacetimes?  Such generalization for a connection that is asymmetric but compatible with the metric structure was performed in Ref. \cite{AMar1991}.  There, the idea was to assume that the field equations derive from an action that is invariant under local Lorentz transformations.  We can follow a more restrictive and, arguable, more physical assumption:  that the action is invariant under both diffeomorphisms and local Poincaré gauge transformations.  Thus, following Refs. \cite{Iosifidis:2023eom,Iosifidis:2021bad,Iosifidis:2021nra,Iosifidis:2020gth}, we will use the ``conservation laws'' in this context:
\begin{align}
\frac{1}{\sqrt{-\det g}} (\nabla_\mu - T_\mu)(\sqrt{-\det g} \Sigma^\mu_{\hspace{2mm}\alpha}) = & -T_{\nu \mu \alpha} \Sigma^{\mu \nu} \nonumber \\
&+ \frac{1}{2} R_{\lambda \mu \nu \alpha} \Delta^{\lambda \mu \nu} \,, \label{cemcl}
\end{align}
and
\begin{equation}
\frac{1}{2\sqrt{-\det g}} (\nabla_\nu - T_\nu) (\sqrt{-\det g} \Delta_\lambda^{\hspace{2mm} \mu \nu}) = \Sigma^\mu_{\hspace{2mm} \lambda} - \tau^\mu_{\hspace{2mm} \lambda} \,, \label{hcl}
\end{equation}
\textcolor{black}{where} 
\begin{equation}
\Delta^\gamma_{\hspace{2mm} \nu \beta} \equiv -\frac{2}{\sqrt{-\det g}} \frac{\delta (\sqrt{-\det g} \mathcal{L}_M)}{\delta \Gamma_{\gamma}{}^{\nu \beta}} \,, \label{hyperm}
\end{equation}
are the components of the hypermomentum tensor.\footnote{\textcolor{black}{The $\nabla$ symbols in Eqs. (\ref{cemcl}) and (\ref{hcl}) are not actually covariant derivatives because they are acting on objects that are not tensors. In practical terms, however, they act like covariant derivatives do so that the $\nabla$ symbol is a very convenient notation.}}

Starting from Eq. (\ref{EC05}), \textcolor{black}{with $\mathcal{T} \equiv \tau$,} raising the index $\alpha$ and applying the operator $\nabla_\alpha$, we obtain
\begin{align}
 \nabla_\alpha \tau^\alpha_{\hspace{2mm} \gamma} =& \frac{1}{8\pi G } \Big\{ \nabla_\gamma f + \nabla_\alpha R^{(\alpha}_{\hspace{3mm} \gamma)} \nonumber \\
 &+ \nabla_\alpha \left[(\nabla_\nu - T_\nu) T^{(\alpha \hspace{4mm} \nu}_{\hspace{4mm} \gamma)} + \nabla^{(\alpha} T_{\gamma)} \right] \Big\} \,. \label{memteq}
\end{align}
Eqs. (\ref{cemcl}), (\ref{hcl}), and (\ref{memteq}) form a system of first-order differential equations for $\Sigma,\Delta$ and $\tau$;
therefore, if there exists a solution in terms of $f,T$ and $g$, it is unique except for an integration constant. The solution to this system of equations exists and requires, in fact, $f = -R/2 + \Lambda-(\nabla_\alpha - T_\alpha/2) T^{\alpha}$ as a consistency condition;  a quick and overwhelming reason is that this $f$ reproduces, via Eq. (\ref{EC05}) \textcolor{black}{with $\mathcal{T} \equiv \tau$}, the gravitational field equation obtained from the action 
\begin{empheq}[box=\fbox]{align}
S&= S_{\rm grav} + S_{M} \nonumber\\
&=
  \frac{1}{16 \pi G}\int d^4 x \sqrt{-\det g}\left[R-2 \Lambda + T_\alpha T^\alpha \right] \nonumber \\
  &+\int d^4 x \sqrt{- \det g}\mathcal{L}_{M} \,,  \label{newtheoryot}
\end{empheq}
which is a slight deviation from the Einstein-Hilbert action, and that, in turn, satisfies Eqs. (\ref{cemcl}) and (\ref{hcl}), it being one of the members of the PGT \cite{blagojevic2001gravitation,Blagojevic:2013xpa,Hehl:1976kj,Hehl1979}.  An interesting property of this action is that it is free from ghosts since the torsion is not a propagating degree of freedom \cite{BeltranJimenez:2019hrm}. 
As in the Riemannian case, the generalization of the other three Lanczos-Lovelock hypotheses:  that the right member of Eq. (\ref{EC05}) i) is built out from the metric tensor, the torsion, and their first- and second-order derivatives, ii) is symmetric, and iii) is linear in second-order derivatives of the metric \textcolor{black}{tensor} and the torsion, are easily seen to be satisfied.\footnote{Since the reason \textcolor{black}{for} the original third hypothesis is that the curvature is the only geometrical descriptor of the gravitational interaction, we could generalize it by requiring the right member of Eq. (\ref{EC05}) to be linear in both the curvature \textcolor{black}{and the torsion. This} would be too radical, though, since it would be impossible to carry out for any function $f$.}

To summarize, the field equation of the {\em unique} gravitational theory (except for the value of the integration constant) selected by Jacobson's thermodynamic approach, \textcolor{black}{when $\mathcal{T} \equiv \tau$,} together with the Lanczos-Lovelock hypotheses is 
\begin{empheq}[box=\fbox]{align}
8\pi G \tau_{\mu \nu} =& R_{(\mu \nu)} -\frac{1}{2}g_{\mu \nu}(R-2\Lambda) +(\nabla_{\alpha}-T_{\alpha})T_{(\mu \nu)}{}{}{}{}^{\alpha}\nonumber \\
&+\nabla_{(\mu}T_{\nu)}- g_{\mu \nu} \left(\nabla_\alpha - \frac{1}{2} T_\alpha \right) T^{\alpha}  \,,
\label{finalJLLT}
\end{empheq}
where $\Lambda$ is a constant.

Finally, it is worthwhile mentioning that Refs. \cite{Dey:2017fld,DeLorenzo:2018odq} claim to have found the EC theory from equilibrium thermodynamics after having implemented a local conservation equation.  We have serious doubts about the logic employed in both papers:  in addition to the concern that raises from footnote \ref{foot6}, the expression for $f$ is claimed in \cite{Dey:2017fld} to have been obtained after having enforced the covariant derivative of Eq. (\ref{finalJLLT}) (their Eq. (96)) which, clearly, is a circular argument; a stronger circular argument is employed in \cite{DeLorenzo:2018odq} where the conserved effective energy-momentum tensor of the EC theory is declared from the beginning of the paper.
Here, in contrast, our arguments follow the strict rules of logic and lead to the gravitational theory described by Eq. (\ref{finalJLLT}).

\subsubsection{\textcolor{black}{$\mathcal{T} \equiv \Sigma$}}

\textcolor{black}{Let us now suppose that $\mathcal{T}$ is identified as the canonical energy-momentum tensor $\Sigma$. As can be demonstrated by employing Eqs. (\ref{hcl}), (\ref{hyperm}), (\ref{eqB06}), and (\ref{eqB11}) (see Ref. \cite{Hehl:1976kj}), the field equation that results from the variation of the Einstein-Hilbert action with respect to the metric tensor acquires the very nice form
\begin{equation}
R_{\mu \nu}-\frac{1}{2}g_{\mu \nu}(R-2\Lambda) = 8\pi G \Sigma_{\mu \nu} \,.
\end{equation}
Comparing the symmetric part of this field equation with Eq. (\ref{EC05}), for $\mathcal{T} \equiv \Sigma$, we can conclude that the EC theory does not belong to the family of gravitational theories selected by Jacobson's thermodynamic approach.}

\textcolor{black}{Following the same logic as in the previous subsubsection, to select a unique gravitational theory from the pool of theories described by the field equation in Eq. (\ref{EC05}), with $\mathcal{T} \equiv \Sigma$, we can assume this theory to be derivable from an action that is invariant under both diffeomorphisms and local Poincaré gauge transformations.  The simultaneous solution to Eqs. (\ref{cemcl}), (\ref{hcl}), and (\ref{memteq}), the latter with $\tau$ exchanged by $\Sigma$, if it exists, will give us such a theory up to an integration constant.  The way to try to find out this solution starts by comparing Eq. (\ref{EC05}), with $\mathcal{T} \equiv \Sigma$, and (\ref{EC06}). The latter is a proxy to find out the action Eq. (\ref{EC05}) derives from as the experiments strongly constrain the presence of torsion.  Thus, since the difference between Eq. (\ref{EC05}), with $\mathcal{T} \equiv \Sigma$, and (\ref{EC06}) is $(\nabla_\sigma - T_\sigma) T_{(\mu \nu)}^{\;\;\;\;\;\;\sigma} + \nabla_{(\mu} T_{\nu)}$, it is clear that, in a first step, the searched action must be of the form 
\begin{equation}
S_{\rm grav} \supset S_{EH} + \frac{1}{16 \pi G}\int d^4 x \sqrt{-\det g} \ (2 \nabla_\sigma - T_\sigma) T^\sigma \,,
\end{equation}
which, however, in unable to generate a term of the form $(\nabla_\sigma - T_\sigma) T_{(\mu \nu)}^{\;\;\;\;\;\;\sigma}$.  The conclusion is that, {\em when $\mathcal{T} \equiv \Sigma$, the field equation obtained from Jacobson's thermodynamic approach does not follow from an action principle}.
}

\textcolor{black}{To summarize, the field equation of the family of gravitational theories selected by Jacobson's thermodynamic approach is given by Eq. (\ref{EC05}) where $f$ is an arbitrary scalar field. The set of Lanczos-Lovelock hypotheses and the Jacobson's procedure are mutually inconsistent when $\mathcal{T} \equiv \Sigma$.}

\section{The Gravitational Fundamental Relation in a General Non-Riemannian Spacetime} \label{sec4}

Up to now, Jacobson's ideas have been an alternative and very interesting tool to obtain different theories of gravity; in particular, GR. Therefore, Jacobson’s procedure could be the key to find the best classical gravitational theory. In this section, we will follow the same strategy of the previous section to derive the generalization of Eq. (\ref{EC05}) to the case where non metricity is also present.

The key point is the integration of Eq. (\ref{RAY10}) leaving only the terms that are linear in $\lambda$:
\begin{widetext}
\begin{align}
\Theta \approx & \lambda \Big[ -R_{\alpha \beta}l^{\alpha}l^{\beta}-l^{\alpha}l^{\beta}\nabla_{\alpha}T_{\beta}+l^{\alpha}l^{\beta}T_{\alpha \beta \nu}T^{\nu}-l^{\alpha}l^{\beta}\nabla_{\nu}T_{\beta \alpha}^{\hspace{4mm}\nu} \nonumber \\
&+\frac{1}{2}l^{\alpha}l^{\beta}\nabla_{\alpha}\tilde{Q}_{\beta}-\frac{1}{2}l^{\alpha}l^{\beta}Q_{\alpha \beta \nu}\tilde{Q}^{\nu} -l^{\alpha}l^{\beta}\nabla_{\nu}Q_{\beta \alpha}{}{}^{\nu}
+l^{\alpha}l^{\beta}T^{\nu}Q_{\alpha \beta \nu}-\frac{1}{2}l^{\alpha}l^{\beta}T_{\alpha \beta \nu}\tilde{Q}^{\nu} \Big] \,.
\end{align}

Thus, by invoking the area law of entropy, $\delta S = \eta \delta \mathcal{A}$, and the Unruh temperature, $T = \hbar \kappa /2\pi$, the Clausius relation $\delta Q = T \delta S$ acquires the form\footnote{\textcolor{black}{It is worthwhile noticing that the identification of the left member of Eq. (\ref{firstlaw?}) with $\delta Q$ via the thermodynamics' first law is still valid if there exist spin and/or shear densities but is not if there exists dilatation density. The complete treatment of the situation under study employing the non-Riemannian generalization of the first law encountered in \cite{Iosifidis:2023kyf}, which involves the dilatation density, goes beyond the scope of this paper and will be treated in a future work.}}
\begin{align}
\kappa\int_{\mathscr{H}}\lambda \textcolor{black}{\mathcal{T}_{\mu \nu}}l^{\mu}l^{\nu}d\lambda d\mathcal{A}=& \frac{\eta \hbar \kappa}{2 \pi}\int_{\mathscr{H}} \lambda \Big[ R_{\alpha \beta}l^{\alpha}l^{\beta}+l^{\alpha}l^{\beta}\nabla_{\alpha}T_{\beta}-l^{\alpha}l^{\beta}T_{\alpha \beta \nu}T^{\nu}+l^{\alpha}l^{\beta}\nabla_{\nu}T_{\beta \alpha}^{\hspace{4mm}\nu} \nonumber \\
& -\frac{1}{2}l^{\alpha}l^{\beta}\nabla_{\alpha}\tilde{Q}_{\beta}+\frac{1}{2}l^{\alpha}l^{\beta}Q_{\alpha \beta \nu}\tilde{Q}^{\nu} +l^{\alpha}l^{\beta}\nabla_{\nu}Q_{\beta \alpha}{}{}^{\nu}
-l^{\alpha}l^{\beta}T^{\nu}Q_{\alpha \beta \nu} +\frac{1}{2}l^{\alpha}l^{\beta}T_{\alpha \beta \nu}\tilde{Q}^{\nu} \Big] \ d\lambda d\mathcal{A} \,. \label{firstlaw?}
\end{align}

Since this expression is valid for all local Rindler horizons, we end up with
\begin{align}
\frac{2\pi}{\hbar \eta} \textcolor{black}{\mathcal{T}_{\alpha \beta}}l^{\alpha}l^{\beta}=& \Big[ R_{\alpha \beta}l^{\alpha}l^{\beta}+l^{\alpha}l^{\beta}\nabla_{\alpha}T_{\beta}-l^{\alpha}l^{\beta}T_{\alpha \beta \nu}T^{\nu}+l^{\alpha}l^{\beta}\nabla_{\nu}T_{\beta \alpha}^{\hspace{4mm}\nu} \nonumber \\
& -\frac{1}{2}l^{\alpha}l^{\beta}\nabla_{\alpha}\tilde{Q}_{\beta}+\frac{1}{2}l^{\alpha}l^{\beta}Q_{\alpha \beta \nu}\tilde{Q}^{\nu} +l^{\alpha}l^{\beta}\nabla_{\nu}Q_{\beta \alpha}{}{}^{\nu}
-l^{\alpha}l^{\beta}T^{\nu}Q_{\alpha \beta \nu} +\frac{1}{2}l^{\alpha}l^{\beta}T_{\alpha \beta \nu}\tilde{Q}^{\nu} \Big] \,,
\end{align}
\end{widetext}
which, in turn, leads to
\begin{align}
\frac{2\pi}{\hbar \eta}\textcolor{black}{\mathcal{T}_{(\alpha \beta)}}=&  R_{(\alpha \beta)}+\nabla_{(\alpha}T_{\beta)}-T_{(\alpha \beta) \nu}T^{\nu}+\nabla_{\nu}T_{(\beta \alpha)}^{\hspace{5mm}\nu} \nonumber \\
& -\frac{1}{2}\nabla_{(\alpha}\tilde{Q}_{\beta)}+\frac{1}{2}Q_{(\alpha \beta) \nu}\tilde{Q}^{\nu} +\nabla_{\nu}Q_{(\beta \alpha)}{}{}^{\nu} \nonumber \\
&-T^{\nu}Q_{(\alpha \beta) \nu} +\frac{1}{2}T_{(\alpha \beta) \nu}\tilde{Q}^{\nu} 
+ f g_{\alpha \beta} \,,
\end{align}
where $f$ is, again, an unknown scalar field. This expression may be written more elegantly as
\begin{empheq}[box=\fbox]{align}
\frac{2\pi}{\hbar \eta}\textcolor{black}{\mathcal{T}_{(\alpha \beta)}} =& R_{(\alpha \beta)}+f g_{\alpha \beta} \nonumber \\
&+\Big[\nabla_{\nu}-T_{\nu}+\frac{1}{2}\tilde{Q}_{\nu}\Big]\left(T_{(\beta \alpha)}^{\hspace{5mm}\nu}+Q_{(\beta \alpha)}^{\hspace{6mm}\nu}\right) \nonumber \\
&+\nabla_{(\alpha} \left(T_{\beta)} - \frac{1}{2} \tilde{Q}_{\beta)}\right) \,. \label{master}
\end{empheq}
Eq. (\ref{master}) identifies the family of gravity theories that are compatible with Jacobson's thermodynamic approach in a general non-Riemannian spacetime and is, in consequence, one of the most important equations in this section.

To select one of the infinite possibilities regarding the scalar field $f$, we will now try to implement the generalization of the Lanczos-Lovelock hypotheses.  There does not exist a generalized version of the Lanczos-Lovelock work in a general non-Riemannian spacetime. Studies in \cite{JimenezCano:2021rlu,Janssen:2019doc,Janssen:2019uao} have been done in this direction but only reaching partial results. Our starting point will be the ``conservation laws'' that arise when requesting the action to be invariant under both diffeomorphisms and local transformations of the affine group  \cite{Iosifidis:2023eom,Iosifidis:2021bad,Iosifidis:2021nra,Iosifidis:2020gth}:
\begin{align}
\frac{1}{\sqrt{-\det g}} (\nabla_\mu - T_\mu)(\sqrt{-\det g} \Sigma^\mu_{\hspace{2mm}\alpha}) = & -T_{\nu \mu \alpha} \Sigma^{\mu \nu} \nonumber \\
&+ \frac{1}{2} R_{\lambda \mu \nu \alpha} \Delta^{\lambda \mu \nu} \nonumber \\
&+ \frac{1}{2} Q_{\alpha \mu \nu} \tau^{\mu \nu} \,, \label{clnr1}
\end{align}
and
\begin{equation}
\frac{1}{2\sqrt{-\det g}} (\nabla_\nu - T_\nu) (\sqrt{-\det g} \Delta_\lambda^{\hspace{2mm} \mu \nu}) = \Sigma^\mu_{\hspace{2mm} \lambda} - \tau^\mu_{\hspace{2mm} \lambda} \,. \label{clnr2}
\end{equation}
Raising the index $\alpha$ and then applying the operator $\nabla_\alpha$ on Eq. (\ref{master}), we obtain
\begin{align}
 \nabla_\alpha \textcolor{black}{\mathcal{T}_{ \hspace{4mm}  \gamma ) }^{ ( \alpha }}  =& \frac{1}{8\pi G} \Big\{ \nabla_\gamma f + \nabla_\alpha R^{(\alpha}_{\hspace{3mm} \gamma)} \nonumber \\
&+\nabla_\alpha \Big[g^{\alpha \lambda}\left(\nabla_\nu - T_\nu+\frac{1}{2}\tilde{Q}_{\nu}\right) \nonumber \\
&\left(T_{(\gamma \lambda)}^{\hspace{5mm}\nu}+Q_{(\gamma \lambda)}^{\hspace{6mm}\nu} \right)\Big] \nonumber \\
&+\nabla_{\alpha}\nabla^{(\alpha} \left(T_{\gamma)} - \frac{1}{2} \tilde{Q}_{\gamma)}\right) \Big\}   \,. \label{clnr3}
\end{align}
Eqs. (\ref{clnr1}), (\ref{clnr2}), and (\ref{clnr3}) form a system of first-order differential equations for $\Sigma,\Delta$ and $\tau$;
therefore, if there exists a solution in terms of $f,T,Q$ and $g$, it is unique except for an integration constant.  The way to try to find out this solution starts by comparing Eq. (\ref{master}) with \textcolor{black}{the symmetric part of the} field equation obtained from the Einstein-Hilbert action (see Eq. (\ref{eqB11})):
\textcolor{black}{
\begin{align}
&R_{(\mu \nu)}-\frac{1}{2}g_{\mu \nu}(R-2\Lambda)+\Big(\nabla_{\alpha}-T_{\alpha}+\frac{1}{2}\tilde{Q}_{\alpha} \Big) \nonumber \\
&\Big[T_{(\nu \mu)}{}{}{}^{\alpha}+Q_{(\nu \mu)}{}{}{}^{\alpha} + \delta^\alpha_{(\nu} \left(T_{\mu)} - \frac{1}{2} \tilde{Q}_{\mu)} \right)
-\frac{1}{2}Q^{\alpha}{}_{\mu \nu} \nonumber \\
&-\frac{1}{2}g_{\mu \nu}\Big(2T^{\alpha}-\tilde{Q}^{\alpha}+Q^{\alpha}\Big)\Big] =8\pi G \tau_{\mu \nu} \,. \label{Qsymfeq}
\end{align}}
The latter is a proxy to find out the action Eq. (\ref{master}) derives from as the experiments strongly constrain the presence of torsion and non metricity.  Thus, it is clear that, in a first step, \textcolor{black}{and when $\mathcal{T} \equiv \tau$,} the searched action must be of the form
\begin{align}
S_{\rm grav} \supset S_{EH} +& \frac{1}{16 \pi G}\int d^4 x \sqrt{-\det g} \nonumber \\
&\Big[-\Big(T^{\alpha}-\frac{1}{2}\tilde{Q}^{\alpha}\Big)\Big(T_{\alpha}-\frac{1}{2}\tilde{Q}_{\alpha}\Big) \nonumber \\
&+ \left(\nabla_\alpha - T_\alpha + \frac{1}{2} \tilde{Q}_\alpha \right)\left(-\frac{1}{2} \tilde{Q}^\alpha \right) \nonumber \\
& - \frac{1}{2} \tilde{Q}_\alpha (2 T^\alpha - \tilde{Q}^\alpha + Q^\alpha) \Big] \,, \label{1stit}
\end{align}
where $S_{EH}$ is the Einstein-Hilbert action plus the cosmological constant term. In a second step, the field equation that derives from the right member in Eq. (\ref{1stit}) is compared with \textcolor{black}{Eq. (\ref{Qsymfeq})} and new terms are added to $S_{\rm grav}$ to try to obtain Eq. (\ref{master}) for a suitable $f$.  The procedure iterates until a match between the two field equations is reached. In fact, this was the procedure we followed to obtain the action in Eq. (\ref{newtheoryot}) in the previous section. Unfortunately, the treatment of the $\nabla_\alpha \tilde{Q}^\alpha$ term in Eq. (\ref{1stit}) requires an infinite number of iterations.  Our conclusion is that Jacobson's approach and the Lanczos-Lovelock hypotheses are mutually inconsistent since Eq. (\ref{master}) does not derive from an action \textcolor{black}{principle}.
Despite this conclusion, and as in the Riemannian case and the non-Riemannian case with vanishing non metricity, the generalization of the other three Lanczos-Lovelock hypotheses:  that the right member of Eq. (\ref{master}) i) is built out from the metric tensor, the torsion, the non metricity, and their first- and second-order derivatives, ii) is symmetric, and iii) is linear in second-order derivatives of the metric, the torsion, and the non metricity, are easily seen to be satisfied.

To summarize, the field equation of the family of gravitational theories selected by Jacobson's thermodynamic approach, \textcolor{black}{when no dilatation density is present,} is given by Eq. (\ref{master}) where $f$ is an arbitrary scalar field. The set of Lanczos-Lovelock hypotheses and the Jacobson's procedure are, in the presence of non metricity, mutually inconsistent. \textcolor{black}{The case when $\mathcal{T} \equiv \Sigma$ will be left for a future work where the dilatation density will be considered.}

\subsection{On the Role of the Non-Equilibrium Contributions to $\delta S$}

The non-equilibrium contribution to $\delta S$ is given by
\begin{align}
\delta S_i=& \eta \int_{\mathscr{H}}\Big(-T_{\mu \rho \nu}\nabla^{\nu}(l^{\mu}l^{\rho})-(\nabla_{\mu}l^{\alpha})(\nabla_{\alpha}l^{\mu}) \nonumber \\
&+T_{\beta \mu \alpha}l^{\alpha}\nabla^{\beta}l^{\mu}
+T^{\nu}l_{\mu}\nabla_{\nu}l^{\mu}-\frac{1}{2}\tilde{Q}_{\nu}l_{\mu}\nabla^{\nu}l^{\mu} \nonumber \\
&-Q_{\rho \mu \nu}\nabla^{\nu}(l^{\mu}l^{\rho}) -\nabla_{\nu}(l_{\mu}\nabla^{\nu}l^{\mu})\Big)\lambda d\lambda d\mathcal{A} \,. \label{MAGneq}
\end{align}
Notice that when one takes vanishing non metricity, Eq. (\ref{MAGneq}) provides the same dissipative terms obtained in \cite{Dey:2017fld} (except for the ambigous term), i.e., those presented in footnote \ref{foot6}. These terms are interpreted by the authors of Ref. \cite{Dey:2017fld} as the possible generalization to the Riemann–Cartan geometries of the Hartle–Hawking term describing the dissipation of a distortion of the horizon. Ref. \cite{Dey:2017fld} also mentions that the new dissipation term could imply a different output of gravitational waves with respect to what is expected in GR when there is matter flowing through the horizons. This key idea could apply to our generalized case, i.e., we might consider Eq. (\ref{MAGneq}) as the generalization to the non-Riemannian geometries of the Hartle-Hawking term. Moreover, the presence of non metricity could imply a different output for gravitational waves with respect to what is expected in GR \textcolor{black}{or the theories described in Eqs. (\ref{newtheoryot}), for $\mathcal{T} \equiv \tau$, and (\ref{EC05}), for $\mathcal{T} \equiv \Sigma$,} when there is matter flowing through the horizons, this flow being also a source of non metricity.

\section{Conclusions} \label{sec7}
Here, we have addressed the idea of generalizing Jacobson's formalism \cite{Jacobson:1995ab} to a general non-Riemannian spacetime. By employing two simple hypotheses: i) that the spacetime is described by a smooth manifold $\mathcal{M}$ with affine and metric structures and ii) that the gravitational interaction has an intimate relation to the thermodynamics' laws, we have found a selected family of gravitational theories. This family includes GR, if $\mathcal{M}$ is Riemannian, but does not include the EC theory, if the affine structure of $\mathcal{M}$ is such that the connection is asymmetric but compatible with the metric structure.   
The number of members in this family is infinite for each of the cases considered, so we have combined Jacobson's thermodynamic approach with the Lanczos-Lovelock hypotheses \cite{Lanczos:1932fxw,Lanczos:1938sf,Lovelock:1969vyr,Lovelock:1970zsf,Lovelock:1971yv,Lovelock:1972vz} in the hope of significantly reducing this number.  The results are particularly appealing:  whereas GR and the theory derivable from the action in Eq. (\ref{newtheoryot}) are the only ones that remain if just curvature or both curvature and torsion are present, \textcolor{black}{with $\mathcal{T} \equiv \tau$,} respectively, there is no way to consistently merge the Jacobson's approach and the Lanczos-Lovelock hypotheses when non metricity shows up, \textcolor{black}{up to the hypotheses that $\mathcal{T} \equiv \tau$ and that no dilatation density is present, or when just curvature and torsion are present and $\mathcal{T} \equiv \Sigma$}. Of course, GR is also derivable from the action in Eq. (\ref{newtheoryot}) in the limit when $T = 0$.  Such an action is a slight deviation from the Einstein-Hilbert action that involves a quadratic term in the torsion vector; we, therefore, can conclude that the Ockham's razor philosophy seems to be operating at a fundamental level and, in consequence, {\em Nature would have selected one of the simplest gravitational actions and would have preferred a connection compatible with the metric structure}.

The thermodynamical argument used here implies that, \textcolor{black}{when $\mathcal{T} \equiv \tau$,} the torsion 
does not introduce new propagating degrees of freedom (i.e., it is algebraically coupled to matter, see Eq. (\ref{newtheoryot})). 
From an observational point of view, this scenario involving non-propagating torsion  
is supported by recent data from multi-messenger astronomy. Additional propagating modes would typically introduce: (i) deviations in the speed of gravitational waves (constrained by GW170817 and GRB 170817A \cite{Baker:2017hug,Creminelli:2017sry,Sakstein:2017xjx,Ezquiaga:2017ekz}), (ii) extra polarization modes, constrained by LIGO/Virgo observations \cite{Callister:2017ocg,LIGOScientific:2018dkp,Hagihara:2018azu, KAGRA:2021kbb}, and (iii) additional energy loss channels, i.e., deviations from pure quadrupolar gravitational radiation, which is tightly constrained by the orbital decay of binary pulsars \cite{Yunes:2010qb,Zhu:2018etc}. 

On the other hand, the presence of non metricity provides an extra contribution to the internal entropy in the full non-Riemannian case so that Eq. (\ref{MAGneq}) could be considered as the generalization of the Hartle-Hawking term to non-Riemannian geometries. The presence of non metricity could imply a different output for gravitational waves, compared to what is expected in GR, \textcolor{black}{the theory derivable from the action in Eq. (\ref{newtheoryot}), or the theory described by the field equation in Eq. (\ref{EC05}) with $\mathcal{T} \equiv \Sigma$, when we have} matter fluxes which are also sources of non metricity at the horizons. This could have possible implications for future observations.

Notice that, even in the non-Riemannian cases, the cosmological constant remains as enigmatic as ever. Neither torsion nor non metricity provides any clues about the nature of $\Lambda$. Thus, this scenario reinforces the idea of interpreting the cosmological constant as the energy contribution from vacuum.

As in previous works \cite{Jacobson:1995ab,Dey:2017fld,DeLorenzo:2018odq}, our result depends on the existence of local equilibrium conditions. Local Rindler horizons, which are instantaneously stationary, are systems that satisfy this requirement. The Clausius relation only applies to variations between nearby states of local thermodynamic equilibrium. Local temperature and entropy are not well defined away from this configuration.\footnote{As is well known, non-equilibrium thermodynamics is required in processes like the free expansion of a gas \textcolor{black}{\cite{callen}}.} When the local equilibrium condition breaks down, the gravitational field would no longer be described \textcolor{black}{by the field equation in Eq. (\ref{EC05}) or Eq. (\ref{master})} and corrections can appear (see, e.g., \cite{Eling:2006aw,Dey:2016zka,Guedens:2011dy}).

One of the key consequences of Jacobson's programme is that it is hinting to an emergent origin for gravity, i.e., that the spacetime dynamics is just a manifestation of the dynamical granular components of the universe itself. Some \textcolor{black}{classical} gravitational theories beyond GR \cite{Eling:2006aw,Dey:2016zka,Guedens:2011dy}, including the ones studied in this paper, have followed this thermodynamic interpretation  of the gravitational interaction, enforcing the idea of gravity as an effective field theory, even in the presence of torsion and non metricity.  \textcolor{black}{We do not know the microscopic description of the gravitational interaction (for some proposals, see, e.g. \cite{Smolin:2012ys,Chirco:2014saa}) but its macroscopic description seems to align with our traditional way of describing this interaction as due to the topological and geometrical dynamics of a classical spacetime.  Recalling H. B. Callen in \cite{callen}, although ``thermodynamics is the study of the restrictions on the possible properties of matter that follow from the symmetry properties of the fundamental laws of physics'' (see, especially, his postlude), there is no reason why the thermodynamic fundamental relation can be derivable from an action principle.  Therefore, it is surprising and remarkable, indeed, that we have been able to find out actions that give way to the fundamental relation for some of the cases considered. What this is telling us about the nature of the gravitational interaction, we have not been able to interpret it yet.}

\begin{acknowledgements}
Jhan N. Martínez has been funded by Universidad Industrial de Santander through the Forgivable Loan Programme. Yeinzon Rodríguez has been funded by Universidad Antonio Nariño under grant numbers VCTI 2024211 and VCTI 2025205. We thank José G. Acevedo for useful discussions \textcolor{black}{and the referees of this paper for their wise advice}.
\end{acknowledgements}

\appendix
\section{Some useful definitions and properties in a general non-Riemannian spacetime}\label{appendix1}

The components of the Riemann or curvature tensor, associated with the general linear connection whose components are $\Gamma^{\mu}{}_{\alpha\beta}$, are given by
\begin{multline}
R^{\rho}{}_{\sigma \mu \nu} \equiv \partial_{\mu}\Gamma^{\rho}{}_{\nu \sigma}-\partial_{\nu}\Gamma^{\rho}{}_{\mu \sigma}
+\Gamma^{\rho}{}_{\mu \lambda}\Gamma^{\lambda}{}_{\nu \sigma}-\Gamma^{\rho}{}_{\nu \lambda}\Gamma^{\lambda}{}_{\mu \sigma} \,.
\label{eqA02}
\end{multline}
These can be split into two parts:
\begin{align}
R^{\rho}{}_{\sigma \mu \nu}=&\partial_{\mu}\mathring{\Gamma}^{\rho}{}_{\nu \sigma}-\partial_{\nu}\mathring{\Gamma}^{\rho}{}_{\mu \sigma}+\mathring{\Gamma}^{\rho}_{\hspace{3mm}\mu \lambda}\mathring{\Gamma}^{\lambda}_{\hspace{3mm}\nu \sigma} \nonumber \\
&-\mathring{\Gamma}^{\rho}_{\hspace{3mm}\nu \lambda}\mathring{\Gamma}^{\lambda}_{\hspace{3mm}\mu \sigma}+\mathring{\nabla}_{\mu}\tilde{\Gamma}^{\rho}_{\hspace{3mm}\nu \sigma}-\mathring{\nabla}_{\nu}\tilde{\Gamma}^{\rho}{}_{\mu \sigma}\nonumber\\
&+\tilde{\Gamma}^{\rho}_{\hspace{3mm}\mu \lambda}\tilde{\Gamma}^{\lambda}_{\hspace{3mm}\nu \sigma}-\tilde{\Gamma}^{\rho}_{\hspace{3mm}\nu \lambda}\tilde{\Gamma}^{\lambda}_{\hspace{3mm}\mu \sigma} \nonumber\\
=&\mathring{R}^{\rho}{}_{\sigma \mu \nu}+\tilde{R}^{\rho}{}_{\sigma \mu \nu} \,,
\label{eqA03}
\end{align}
where $\mathring{R}^{\rho}{}_{\sigma \mu \nu}$ are the components of the Riemann tensor associated with the Levi-Civita connection and $\tilde{R}^{\rho}{}_{\sigma \mu \nu}$ is the contribution from the torsion and non metricity. The components of the Ricci tensor are defined as $R_{\mu \nu}\equiv R^{\alpha}_{\hspace{2mm}\mu \alpha \nu}$ and the Ricci scalar as $R \equiv R_{\mu \nu}g^{\mu \nu}$. Notice that the commutator of the covariant derivatives in a general non-Riemannian spacetime is given by
\begin{equation}
\nabla_{\alpha}\nabla_{\mu}u^{\nu}-\nabla_{\mu}\nabla_{\alpha}u^{\nu}=R^{\nu}{}_{\beta \alpha \mu}u^{\beta}+T^{\beta}{}_{\mu \alpha }\nabla_{\beta}u^{\nu} \,.
\label{eqA04}
\end{equation}

The Riemann tensor is antisymmetric in the last two indices: 
\begin{equation}
R^{\rho}_{\hspace{2mm}\sigma \mu \nu}=-R^{\rho}_{\hspace{2mm}\sigma \nu \mu} \,.
\label{eqA06}
\end{equation}
Other standard symmetries and the Bianchi identities of the Riemann tensor are modified in the presence of torsion and non metricity. For instance, the existence of non metricity implies the following three identities \cite{schouten2013ricci}: the non-antisymmetry identity of the Riemann tensor in its first two indices,
\begin{equation}
R_{\alpha \beta \nu \mu} = -R_{\beta \alpha \nu \mu} + \nabla_{\mu}Q_{\nu \alpha \beta}-\nabla_{\nu}Q_{\mu \alpha \beta}+T^{\delta}_{\hspace{2mm}\mu \nu}Q_{\delta \alpha \beta} \,,
\label{eqA07}
\end{equation}
the first Bianchi identity,
\begin{align}
R^{\rho}_{\hspace{2mm}\sigma \mu \nu}+R^{\rho}_{\hspace{2mm}\mu \nu \sigma}+R^{\rho}_{\hspace{2mm}\nu \sigma \mu}= &\nabla_{\sigma}T^{\rho}_{\hspace{2mm}\mu \nu}+\nabla_{\mu}T^{\rho}_{\hspace{2mm}\nu \sigma} \nonumber \\
&+\nabla_{\nu}T^{\rho}_{\hspace{2mm}\sigma \mu} +T^{\rho}_{\hspace{2mm}\lambda \sigma}T^{\lambda}_{\hspace{2mm}\mu \nu} \nonumber \\
&+T^{\rho}_{\hspace{2mm}\lambda \mu}T^{\lambda}_{\hspace{2mm}\nu \sigma}+T^{\rho}_{\hspace{2mm}\lambda \nu}T^{\lambda}_{\hspace{2mm}\sigma \mu} \,,
\label{eqA08}
\end{align}
from which it follows that
\begin{align}
R_{\sigma \nu} =& R_{\nu \sigma} + \nabla_{\mu}T^{\mu}_{\hspace{2mm}\nu \sigma}+\nabla_{\nu}T^{\mu}_{\hspace{2mm} \sigma \mu}-\nabla_{\sigma}T^{\mu}_{\hspace{2mm} \nu \mu}\nonumber\\
&-T^{\mu}_{\hspace{2mm} \mu \lambda}T^{\lambda}_{\hspace{2mm} \nu \sigma}+\frac{1}{2}g^{\alpha \beta}\nabla_{\nu}Q_{\sigma \alpha \beta}\nonumber\\
&-\frac{1}{2}g^{\alpha \beta}\nabla_{\sigma}Q_{\nu \alpha \beta}\nonumber+\frac{1}{2}g^{\alpha \beta}T^{\delta}_{\hspace{2mm} \nu \sigma}Q_{\delta \alpha \beta}\nonumber\\
=& R_{\nu \sigma} + \Big(\nabla_{\mu}-T_{\mu}+\frac{1}{2} \tilde{Q}_{\mu}\Big)\Big(T^{\mu}_{\hspace{2mm}\nu \sigma}+\delta^{\mu}_{\sigma}T_{\nu} \nonumber \\
&-\delta^{\mu}_{\nu}T_{\sigma}+\frac{1}{2}\delta^{\mu}_{\nu}\tilde{Q}_{\sigma} -\frac{1}{2}\delta^{\mu}_{\sigma}\tilde{Q}_{\nu} \Big) \nonumber\\
=& R_{\nu \sigma} + \Big(\nabla_{\mu}-\tilde{\Gamma}^{\lambda}{}_{\lambda \mu}\Big)\Big(\tilde{\Gamma}^{\mu}{}_{\nu \sigma}-\tilde{\Gamma}^{\mu}{}_{\sigma \nu} \nonumber \\
&+\delta^{\mu}_{\sigma}\tilde{\Gamma}^{\alpha}{}_{\alpha \nu}-\delta^{\mu}_{\nu}\tilde{\Gamma}^{\alpha}{}_{\alpha \sigma} \Big) \,,
\label{eqA09}
\end{align}
and the second Bianchi identity,
\begin{align}
\nabla_{\alpha}R^{\rho}{}_{\sigma \mu \nu}+\nabla_{\mu}R^{\rho}_{\hspace{2mm}\sigma \nu \alpha}+\nabla_{\nu}R^{\rho}_{\hspace{2mm}\sigma \alpha \mu }=& R^{\rho}_{\hspace{2mm}\sigma \lambda \nu}T^{\lambda}_{\hspace{2mm}\mu \alpha} \nonumber \\
&+R^{\rho}_{\hspace{2mm}\sigma \lambda \mu}T^{\lambda}_{\hspace{2mm} \alpha \nu} \nonumber \\
&+R^{\rho}_{\hspace{2mm}\sigma \lambda \alpha}T^{\lambda}_{\hspace{2mm}\nu \mu} \,,
\label{eqA10}
\end{align}
which leads to
\begin{align}
\nabla_{\beta}R^{\beta}{}_{\nu}=&\frac{1}{2}\nabla_{\nu}R+\frac{1}{2}R_{\mu \alpha \lambda \nu}T^{\lambda \mu \alpha}+R^{\mu \lambda}T_{\lambda \nu \mu} \nonumber \\
&-\frac{1}{2}R_{\beta \nu}Q^{\beta}
+\frac{1}{2}R^{\sigma\alpha}Q_{\nu \sigma \alpha}-\frac{1}{2}R_{\mu \sigma \nu \alpha}Q^{\mu \sigma \alpha} \nonumber \\
&-\frac{1}{2}\nabla_{\mu}\Big(g^{\sigma \alpha}g^{\mu \beta} \Big[ \nabla_{\alpha}Q_{\nu \beta \sigma}-\nabla_{\nu}Q_{\alpha \beta \sigma} \nonumber \\
&+T^{\delta}_{\hspace{2mm}\alpha \nu}Q_{\delta \beta \sigma} \Big]\Big) \nonumber \\
&-\frac{1}{2}g^{\alpha \sigma}T^{\lambda \mu}_{\hspace{4mm}\nu}\Big(\nabla_{\alpha}Q_{\lambda \mu \sigma}-\nabla_{\lambda}Q_{\alpha \mu \sigma} \nonumber \\
&+T^{\delta}_{\hspace{2mm}\alpha \lambda}Q_{\delta \mu \sigma} \Big) \,.
\label{eqA11}
\end{align}

\section{Proportionality between $h$ and $g$} \label{fg}

This appendix examines a crucial step in deriving the field equations from the Jacobson's thermodynamic approach. In the steps leading to the field equations, a symmetric tensor $h$ is considered such that $h_{\alpha\beta}l^\alpha l^\beta=0$ for {\em any} null vector $l$.  While previous works assert that $h$ must be proportional to the metric $g$, we provide an explicit demonstration here. A proof of this proportionality exists in \cite{wald2010general} (see Appendix D), but it assumes that both tensors are symmetric and non degenerate. In this appendix, we relax the second condition for $h$ requiring only that it is symmetric, and show that the result remains valid regardless of the presence of curvature, torsion, or non metricity.

There always exists a transformation of the coordinates that puts the components of the metric in canonical form at any point $\mathcal{P}$: $g_{\alpha\beta}|_{\mathcal{P}}=\eta_{\alpha \beta}$ where $\eta_{\alpha \beta}$ are the components of the Minkowski matrix.  We will work in these coordinates.

Since $h_{\alpha\beta}l^\alpha l^\beta=0$ must be valid for \emph{any} null vector, in particular for the six null vectors whose components are $l^0_A=1, l^i_A=\pm\delta_A^i$, with the index $A = 1,2,3$, we can construct the equations
\begin{equation}
    h_{00}\pm 2h_{0i}+h_{ii} =0 \,, \hspace{5mm} i = 1,2,3 \,,  
\end{equation}
where repeated indices do not imply summation. The solution for this system of equations is $h_{ii} = -h_{00}, h_{0i}=0$. 

Let us now choose another vector, for example, $l^\mu_7 = (1, 2^{-1/2},2^{-1/2},0)$. The condition on $h_{\alpha\beta}$ gives in this case:
\begin{equation}
    2 h_{00}+ 2 \sqrt{2} h_{01}+ 2 \sqrt{2} h_{02}+h_{11}+ 2 f_{12}+f_{22} = 0 \,,
\end{equation}
which, given the previous solutions, implies that $h_{12}=0$. Similarly, if we choose $l^\mu_8=(1,2^{-1/2},0,2^{-1/2})$, we obtain $h_{13}=0$, and if we choose $l^\mu_9=(1,0,2^{-1/2},2^{-1/2})$, we obtain $h_{23}=0$. Consequently, 
\begin{equation}
    h_{\alpha\beta} = f\, \eta_{\alpha\beta} \,,
\end{equation}
with $f$ being an arbitrary scalar. This was proven for nine particular null vectors but, given that $h$ turned out to be proportional to the metric, the expression $h_{\alpha\beta}l^\alpha l^\beta=0$ holds for any other null vector. Finally, by transforming back to a general coordinate system, $h_{\alpha\beta}=f\, g_{\alpha\beta}$.

\section{Gravitational field equations from the Einstein-Hilbert action} \label{appendix2}

We consider as independent variables the metric $g$, the torsion $T$ and the non-metricity $Q$. The field equations are obtained by invoking the stationary action principle: 
\begin{equation}
\frac{\delta S}{\delta g^{\mu\nu}} = 0, \quad \frac{\delta S}{\delta T^{\alpha}{}_{\mu\nu}} = 0,\quad \frac{\delta S}{\delta Q_{\alpha}{}^{\mu\nu}} = 0 \,.
\end{equation}

The Einstein-Hilbert action, supplemented with the cosmological constant term and the matter action, is given by
\begin{align}
S&= S_{\rm grav} + S_{M} \nonumber\\
&=
  \frac{1}{16 \pi G}\int d^4 x \sqrt{-\det g}\left(R-2 \Lambda \right) \nonumber \\
  &+\int d^4 x \sqrt{- \det g}\mathcal{L}_{M} \,.
\end{align}
Varying $S_{\rm grav}$ in the previous expression leads to
\begin{align}
16\pi G \ \delta S_{
\rm grav
}=& \int d^4x \sqrt{-\det g}\Big[ R_{\beta \nu}\delta g^{\nu \beta}+g^{\nu \beta}\delta R_{\beta \nu} \nonumber \\
&-\frac{1}{2}g_{\beta \nu}(R-2\Lambda)\delta g^{\nu \beta} \Big] \,.
\label{eqB02}
\end{align}
The explicit variation of the Ricci tensor is given by
\begin{align}
\begin{aligned}
\delta R_{\sigma \nu}&=\nabla_{\mu} \Big(\delta \Gamma^{\mu}{}_{\nu \sigma}  \Big)-\nabla_{\nu} \Big(\delta \Gamma^{\mu}{}_{\mu \sigma}\Big)+T^{\lambda}{}_{\mu \nu}\delta\Gamma^{\mu}{}_{\lambda \sigma}\\
&=\mathring{\nabla}_{\mu} (\delta \Gamma^{\mu}{}_{\nu \sigma})-\mathring{\nabla}_{\nu}(\delta \Gamma^{\mu}{}_{\mu \sigma})+\Big(\delta_{\lambda}^{\alpha}\tilde{\Gamma}^{\mu}{}_{\nu \sigma}\\
&-\delta_{\nu}^{\alpha}\tilde{\Gamma}^{\mu}{}_{\lambda \sigma}-\delta_{\sigma}^{\mu}\tilde{\Gamma}^{\alpha}{}_{\nu \lambda}+\delta^{\alpha}_{\nu}\delta_{\sigma}^{\mu}\tilde{\Gamma}^{\beta}{}_{\beta \lambda} \Big) \delta \Gamma^{\lambda}{}_{\alpha \mu} \,.
\end{aligned}
\label{eqB03}
\end{align}
Then, Eq. \eqref{eqB02} becomes
\begin{align}
&16\pi G \ \delta S_{\rm grav
}=\int d^4x \sqrt{-\det g}\Big[ R_{\beta \nu}\delta g^{\nu \beta} \nonumber \\
&-\frac{1}{2}g_{\beta \nu}(R-2\Lambda)\delta g^{\nu \beta} \nonumber \\
&+g^{\nu \sigma}\Big(\delta_{\lambda}^{\alpha}\tilde{\Gamma}^{\mu}{}_{\nu \sigma}-\delta_{\nu}^{\alpha}\tilde{\Gamma}^{\mu}{}_{\lambda \sigma}-\delta_{\sigma}^{\mu}\tilde{\Gamma}^{\alpha}{}_{\nu \lambda}+\delta^{\alpha}_{\nu}\delta_{\sigma}^{\mu}\tilde{\Gamma}^{\beta}{}_{\beta \lambda} \Big) \delta \Gamma^{\lambda}{}_{\alpha \mu} \Big] \nonumber \\
&+\int d^4x \sqrt{-\det g}\Big[ \mathring{\nabla}_{\mu} (g^{\nu \sigma}\delta \Gamma^{\mu}{}_{\nu \sigma})-\mathring{\nabla}_{\nu}(g^{\nu \sigma}\delta \Gamma^{\mu}{}_{\mu \sigma})\Big] \,.
\label{eqB06}
\end{align}
It is convenient to express the variation of the Levi-Civita connection in terms of the covariant derivative of the variation of the metric as follows \cite{Guarnizo:2010xr}:
\begin{equation}
\delta\mathring{\Gamma}^{\lambda}{}_{\alpha \mu}=\frac{1}{2}g^{\lambda \gamma}\Big( \mathring{\nabla}_{\alpha}\delta g_{\mu \gamma}+\mathring{\nabla}_{\mu}\delta g_{\gamma \alpha}-\mathring{\nabla}_{\gamma}\delta g_{\alpha \mu} \Big) \,.
\label{eqB04}
\end{equation}
Consequently, the variation of the connection is given by \cite{Medina:2018rnl,ortin2004gravity} 
\begin{align}
\begin{aligned}
\delta\Gamma^{\lambda}{}_{\alpha \mu}=&\frac{1}{2}g^{\lambda \gamma}\Big( \nabla_{\alpha}\delta g_{\mu \gamma}+\nabla_{\mu}\delta g_{\gamma \alpha}-\nabla_{\gamma}\delta g_{\alpha \mu} \Big)\\
&+\frac{1}{2}\Big(\delta T^{\lambda}{}_{\alpha \mu}-g^{\lambda \gamma}g_{\mu \beta}\delta T^{\beta}{}_{\alpha \gamma}-g^{\lambda \gamma}g_{\alpha \beta}\delta T^{\beta}{}_{\mu \gamma} \Big)\\
&+\frac{1}{2}g^{\lambda \gamma}\Big(\delta Q_{\gamma \alpha \mu}-\delta Q_{\mu \alpha \gamma}-\delta Q_{\alpha \mu \gamma} \Big) \,.
\end{aligned}
\label{eqB05}
\end{align}

The previous expressions \textcolor{black}{give} us the field equation coming from the variation of the torsion: 
\begin{align}
& \frac{1}{2}\Big(\delta_{\lambda}^{\alpha}\tilde{\Gamma}^{\mu \theta}{}{}_{\theta}-\tilde{\Gamma}^{\mu}{}_{\lambda}{}^{\alpha}-\tilde{\Gamma}^{\alpha \mu}{}{}_{\lambda}+g^{\alpha \mu}\tilde{\Gamma}^{\theta}{}_{\theta \lambda} \Big)\Big(\delta^{\lambda}_{\beta}\delta^{\nu}_{\alpha}\delta^{\gamma}_{\mu} \nonumber \\
& -g^{\lambda \gamma}g_{\mu \beta}\delta^{\nu}_{\alpha}-g^{\lambda \gamma}g_{\alpha\beta}\delta^{\nu}_{\mu} \Big)+16\pi G \frac{\delta \mathcal{L}_M}{\delta T^{\beta}{}_{\nu \gamma}}=0 \,,
\label{eqB07}
\end{align}
and the field equation coming from the variation of the non metricity: 
\begin{align}
&\frac{1}{2}\Big(\delta_{\lambda}^{\alpha}\tilde{\Gamma}^{\mu \theta}{}_{\theta}-\tilde{\Gamma}^{\mu}{}_{\lambda}{}^{\alpha}-\tilde{\Gamma}^{\alpha \mu}{}{}_{\lambda}+g^{\alpha \mu}\tilde{\Gamma}^{\theta}{}_{\theta \lambda} \Big)\Big(g^{\lambda \gamma}g_{\alpha \beta}g_{\mu \nu} \nonumber \\
&-g_{\alpha \beta}\delta^{\lambda}_{\nu}\delta^{\gamma}_{\mu}-g_{\mu \nu}\delta^{\gamma}_{\alpha}\delta_{\beta}^{\lambda} \Big)+ 16\pi G \frac{\delta \mathcal{L}_M}{\delta Q_{\gamma}{}^{\nu \beta}} = 0 \,.
\label{eqB08}
\end{align}
Finally, the variation of $S_{\rm grav}$ with respect to the metric tensor leads to
\begin{align}
16\pi G \ \frac{\delta S_{\rm grav}}{\delta g^{\mu\nu}}\delta g^{\mu\nu}=&\int d^4x \sqrt{-\det g}\Big[  R_{\beta \nu}\delta g^{\nu \beta} \nonumber \\
&-\frac{1}{2}g_{\beta \nu} (R-2\Lambda) \delta g^{\nu \beta} \nonumber \\
& +\mathring{\nabla}_{\mu} (g^{\nu \sigma}\delta \Gamma^{\mu}{}_{\nu \sigma})-\mathring{\nabla}_{\nu}(g^{\nu \sigma}\delta \Gamma^{\mu}{}_{\mu \sigma}) \nonumber \\
&+\frac{1}{2}\Big(\delta_{\lambda}^{\alpha}\tilde{\Gamma}^{\mu \theta}{}{}_{\theta}-\tilde{\Gamma}^{\mu}{}_{\lambda}{}^{\alpha} \nonumber -\tilde{\Gamma}^{\alpha \mu}{}_{\lambda} \nonumber \\
&+g^{\alpha \mu}\tilde{\Gamma}^{\theta}{}_{\theta \lambda} \Big) \Big(g^{\lambda\gamma}g_{\alpha \beta}g_{\mu \nu}\nabla_{\gamma}\delta g^{\nu \beta} \nonumber \\
&-g_{\alpha \beta}\nabla_{\mu}\delta g^{\lambda \beta} -g_{\mu \nu}\nabla_{\alpha}\delta g^{\nu \lambda}\Big) \Big] \,,
\label{eqB09}
\end{align}
and, therefore,
\begin{align}
16\pi G \ \frac{\delta S_{\rm grav}}{\delta g^{\mu\nu}}\delta g^{\mu\nu}
=&\int d^4x \sqrt{-\det g}\Big[  R_{\beta \nu}\delta g^{\nu \beta} \nonumber \\
&-\frac{1}{2}g_{\beta \nu} (R-2\Lambda)\delta g^{\nu \beta} \nonumber \\
&+\mathring{\nabla}_{\mu} (g^{\nu \sigma}\delta \Gamma^{\mu}{}_{
\nu \sigma})-\mathring{\nabla}_{\nu}(g^{\nu \sigma}\delta \Gamma^{\mu}{}_{\mu \sigma}) \nonumber \\
&+\frac{1}{2}\Big(2\tilde{\Gamma}_{[\nu \beta]}{}{}{}^{\gamma}-2\delta^{\gamma}_{\nu}\tilde{\Gamma}^{\theta}{}_{\theta \beta}+2\tilde{\Gamma}^{\gamma}{}_{(\beta \nu)} \nonumber \\
&+g_{\beta \nu}\big[\tilde{\Gamma}_{\theta}{}^{\theta \gamma}-\tilde{\Gamma}^{\gamma \theta}{}{}_{\theta} \big] \Big)\nabla_{\gamma}\delta g^{\nu \beta} \Big] \nonumber \\
=&\int d^4x \sqrt{-\det g}\Big[  R_{\beta \nu}\delta g^{\nu \beta} \nonumber \\
&-\frac{1}{2}g_{\beta \nu} (R-2\Lambda)\delta g^{\nu \beta} \nonumber \\
&+\mathring{\nabla}_{\mu} (g^{\nu \sigma}\delta \Gamma^{\mu}{}_{\nu \sigma})-\mathring{\nabla}_{\nu}(g^{\nu \sigma}\delta \Gamma^{\mu}{}_{\mu \sigma}) \nonumber \\
&+\frac{1}{2}\Big( T^{\gamma}{}_{\nu \beta}-2T_{(\nu \beta)}{}{}{}^{\gamma}+Q^{\gamma}{}_{\nu \beta} \nonumber \\
&-2Q_{\beta \nu}{}{}^{\gamma}-2\delta^{\gamma}_{\nu}\Big(T_{\beta}-\frac{1}{2}\tilde{Q}_{\beta}\Big) \nonumber \\
&+g_{\beta \nu}\big[2T^{\gamma}-\tilde{Q}^{\gamma}+Q^{\gamma} \big] \Big)\nabla_{\gamma}\delta g^{\nu \beta} \Big] \,,
\label{eqB10}
\end{align}
where the squared brackets enclosing indices mean normalized antisymmetrization.
Thus, by integrating by parts, we obtain the field equation coming from the variation of the metric:
\textcolor{black}{\begin{align}
&R_{\mu \nu}-\frac{1}{2}g_{\mu \nu}(R-2\Lambda)+\Big(\nabla_{\alpha}-T_{\alpha}+\frac{1}{2}\tilde{Q}_{\alpha} \Big) \nonumber \\
&\Big[T_{(\nu \mu)}{}{}{}^{\alpha}+Q_{(\nu \mu)}{}{}{}^{\alpha} + \delta^\alpha_{\nu} \left(T_{\mu} - \frac{1}{2} \tilde{Q}_{\mu} \right) -\frac{1}{2}T^{\alpha}{}_{\nu \mu} -\frac{1}{2}Q^{\alpha}{}_{\nu \mu} \nonumber \\
&-\frac{1}{2}g_{\mu \nu}\Big(2T^{\alpha}-\tilde{Q}^{\alpha}+Q^{\alpha}\Big)\Big] =8\pi G \tau_{\mu \nu} \,.
\label{eqB11}
\end{align}}
It is worthwhile noticing that the remaining surface integral is given by
\begin{align}
16\pi G \ \delta S_{
\rm surf 
}
=&\int d^4x \sqrt{-\det g}\Big[\mathring{\nabla}_{\mu} (g^{\nu \sigma}\delta \Gamma^{\mu}{}_{\nu \sigma}) \nonumber \\
&-\mathring{\nabla}_{\nu}(g^{\nu \sigma}\delta \Gamma^{\mu}{}_{\mu \sigma})
+\frac{1}{2}\mathring{\nabla}_{\gamma}\Big( \Big[ T^{\gamma}{}_{\nu \beta} \nonumber \\
&-2 T_{(\nu \beta)}{}{}{}{}{}^{\gamma}+Q^{\gamma}{}_{\nu \beta}-2Q_{\beta \nu}{}{}^{\gamma} \nonumber \\
&-2\delta^{\gamma}_{\nu}\Big(T_{\beta}-\frac{1}{2}\tilde{Q}_{\beta}\Big) \nonumber \\
&+g_{\beta \nu}\Big(2T^{\gamma}-\tilde{Q}^{\gamma}+Q^{\gamma} \Big) \Big]\delta g^{\nu \beta} \Big)\Big] \,.
\label{eqB12}
\end{align}

\bibliography{apssamp,sample}

\providecommand{\noopsort}[1]{}\providecommand{\singleletter}[1]{#1}%
\begin{thebibliography}{80}%
\makeatletter
\providecommand \@ifxundefined [1]{%
 \@ifx{#1\undefined}
}%
\providecommand \@ifnum [1]{%
 \ifnum #1\expandafter \@firstoftwo
 \else \expandafter \@secondoftwo
 \fi
}%
\providecommand \@ifx [1]{%
 \ifx #1\expandafter \@firstoftwo
 \else \expandafter \@secondoftwo
 \fi
}%
\providecommand \natexlab [1]{#1}%
\providecommand \enquote  [1]{``#1''}%
\providecommand \bibnamefont  [1]{#1}%
\providecommand \bibfnamefont [1]{#1}%
\providecommand \citenamefont [1]{#1}%
\providecommand \href@noop [0]{\@secondoftwo}%
\providecommand \href [0]{\begingroup \@sanitize@url \@href}%
\providecommand \@href[1]{\@@startlink{#1}\@@href}%
\providecommand \@@href[1]{\endgroup#1\@@endlink}%
\providecommand \@sanitize@url [0]{\catcode `\\12\catcode `\$12\catcode
  `\&12\catcode `\#12\catcode `\^12\catcode `\_12\catcode `\%12\relax}%
\providecommand \@@startlink[1]{}%
\providecommand \@@endlink[0]{}%
\providecommand \url  [0]{\begingroup\@sanitize@url \@url }%
\providecommand \@url [1]{\endgroup\@href {#1}{\urlprefix }}%
\providecommand \urlprefix  [0]{URL }%
\providecommand \Eprint [0]{\href }%
\providecommand \doibase [0]{https://doi.org/}%
\providecommand \selectlanguage [0]{\@gobble}%
\providecommand \bibinfo  [0]{\@secondoftwo}%
\providecommand \bibfield  [0]{\@secondoftwo}%
\providecommand \translation [1]{[#1]}%
\providecommand \BibitemOpen [0]{}%
\providecommand \bibitemStop [0]{}%
\providecommand \bibitemNoStop [0]{.\EOS\space}%
\providecommand \EOS [0]{\spacefactor3000\relax}%
\providecommand \BibitemShut  [1]{\csname bibitem#1\endcsname}%
\let\auto@bib@innerbib\@empty
\bibitem [{\citenamefont {Faber}(2001)}]{Faber:1999ia}%
  \BibitemOpen
  \bibfield  {author} {\bibinfo {author} {\bibfnamefont {M.}~\bibnamefont
  {Faber}},\ }\bibfield  {title} {\bibinfo {title} {{A Model for topological
  fermions}},\ }\href {https://doi.org/10.1007/s006010170009} {\bibfield
  {journal} {\bibinfo  {journal} {Few Body Syst.}\ }\textbf {\bibinfo {volume}
  {30}},\ \bibinfo {pages} {149} (\bibinfo {year} {2001})},\ \Eprint
  {https://arxiv.org/abs/hep-th/9910221} {arXiv:hep-th/9910221} \BibitemShut
  {NoStop}%
\bibitem [{\citenamefont {Meinert}\ and\ \citenamefont
  {Hofmann}(2024)}]{Meinert:2024lul}%
  \BibitemOpen
  \bibfield  {author} {\bibinfo {author} {\bibfnamefont {J.}~\bibnamefont
  {Meinert}}\ and\ \bibinfo {author} {\bibfnamefont {R.}~\bibnamefont
  {Hofmann}},\ }\bibfield  {title} {\bibinfo {title} {{Electroweak Parameters
  from Mixed SU(2) Yang{\textendash}Mills Thermodynamics}},\ }\href
  {https://doi.org/10.3390/sym16121587} {\bibfield  {journal} {\bibinfo
  {journal} {Symmetry}\ }\textbf {\bibinfo {volume} {16}},\ \bibinfo {pages}
  {1587} (\bibinfo {year} {2024})},\ \Eprint {https://arxiv.org/abs/2401.03243}
  {arXiv:2401.03243 [hep-th]} \BibitemShut {NoStop}%
\bibitem [{\citenamefont {Saikumar}(2024)}]{Saikumar:2024ahz}%
  \BibitemOpen
  \bibfield  {author} {\bibinfo {author} {\bibfnamefont {D.}~\bibnamefont
  {Saikumar}},\ }\bibfield  {title} {\bibinfo {title} {{Exploring the
  Frontiers: Challenges and Theories Beyond the Standard Model}},\ }\href@noop
  {} {\  (\bibinfo {year} {2024})},\ \Eprint {https://arxiv.org/abs/2404.03666}
  {arXiv:2404.03666 [hep-ph]} \BibitemShut {NoStop}%
\bibitem [{\citenamefont {Meinert}\ and\ \citenamefont
  {Hofmann}(2021)}]{Meinert:2021gpb}%
  \BibitemOpen
  \bibfield  {author} {\bibinfo {author} {\bibfnamefont {J.}~\bibnamefont
  {Meinert}}\ and\ \bibinfo {author} {\bibfnamefont {R.}~\bibnamefont
  {Hofmann}},\ }\bibfield  {title} {\bibinfo {title} {{Axial Anomaly in
  Galaxies and the Dark Universe}},\ }\href
  {https://doi.org/10.3390/universe7060198} {\bibfield  {journal} {\bibinfo
  {journal} {Universe}\ }\textbf {\bibinfo {volume} {7}},\ \bibinfo {pages}
  {198} (\bibinfo {year} {2021})},\ \Eprint {https://arxiv.org/abs/2106.01457}
  {arXiv:2106.01457 [hep-ph]} \BibitemShut {NoStop}%
\bibitem [{\citenamefont {Hawking}\ and\ \citenamefont
  {Ellis}(1973)}]{Hawking:1973uf}%
  \BibitemOpen
  \bibfield  {author} {\bibinfo {author} {\bibfnamefont {S.~W.}\ \bibnamefont
  {Hawking}}\ and\ \bibinfo {author} {\bibfnamefont {G.~F.~R.}\ \bibnamefont
  {Ellis}},\ }\href {https://doi.org/10.1017/9781009253161} {\emph {\bibinfo
  {title} {{The Large Scale Structure of Space-Time}}}},\ Cambridge Monographs
  on Mathematical Physics\ (\bibinfo  {publisher} {Cambridge University
  Press},\ \bibinfo {year} {1973})\BibitemShut {NoStop}%
\bibitem [{\citenamefont {Burgess}(2020)}]{Burgess:2020tbq}%
  \BibitemOpen
  \bibfield  {author} {\bibinfo {author} {\bibfnamefont {C.~P.}\ \bibnamefont
  {Burgess}},\ }\href {https://doi.org/10.1017/9781139048040} {\emph {\bibinfo
  {title} {{Introduction to Effective Field Theory}}}}\ (\bibinfo  {publisher}
  {Cambridge University Press},\ \bibinfo {year} {2020})\BibitemShut {NoStop}%
\bibitem [{\citenamefont {Burgess}(2004)}]{Burgess:2003jk}%
  \BibitemOpen
  \bibfield  {author} {\bibinfo {author} {\bibfnamefont {C.~P.}\ \bibnamefont
  {Burgess}},\ }\bibfield  {title} {\bibinfo {title} {{Quantum gravity in
  everyday life: General relativity as an effective field theory}},\ }\href
  {https://doi.org/10.12942/lrr-2004-5} {\bibfield  {journal} {\bibinfo
  {journal} {Living Rev. Rel.}\ }\textbf {\bibinfo {volume} {7}},\ \bibinfo
  {pages} {5} (\bibinfo {year} {2004})},\ \Eprint
  {https://arxiv.org/abs/gr-qc/0311082} {arXiv:gr-qc/0311082} \BibitemShut
  {NoStop}%
\bibitem [{\citenamefont {Lee}(2013)}]{lee2003introduction}%
  \BibitemOpen
  \bibfield  {author} {\bibinfo {author} {\bibfnamefont {J.~M.}\ \bibnamefont
  {Lee}},\ }\href {https://doi.org/10.1007/978-1-4419-9982-5} {\emph {\bibinfo
  {title} {Introduction to Smooth Manifolds}}},\ Graduate Texts in Mathematics\
  (\bibinfo  {publisher} {Springer},\ \bibinfo {year} {2013})\BibitemShut
  {NoStop}%
\bibitem [{\citenamefont {Jacobson}\ and\ \citenamefont
  {Parentani}(2003)}]{Jacobson:2003wv}%
  \BibitemOpen
  \bibfield  {author} {\bibinfo {author} {\bibfnamefont {T.}~\bibnamefont
  {Jacobson}}\ and\ \bibinfo {author} {\bibfnamefont {R.}~\bibnamefont
  {Parentani}},\ }\bibfield  {title} {\bibinfo {title} {{Horizon entropy}},\
  }\href {https://doi.org/10.1023/A:1023785123428} {\bibfield  {journal}
  {\bibinfo  {journal} {Found. Phys.}\ }\textbf {\bibinfo {volume} {33}},\
  \bibinfo {pages} {323} (\bibinfo {year} {2003})},\ \Eprint
  {https://arxiv.org/abs/gr-qc/0302099} {arXiv:gr-qc/0302099} \BibitemShut
  {NoStop}%
\bibitem [{\citenamefont {Jacobson}(1995)}]{Jacobson:1995ab}%
  \BibitemOpen
  \bibfield  {author} {\bibinfo {author} {\bibfnamefont {T.}~\bibnamefont
  {Jacobson}},\ }\bibfield  {title} {\bibinfo {title} {{Thermodynamics of
  space-time: The Einstein equation of state}},\ }\href
  {https://doi.org/10.1103/PhysRevLett.75.1260} {\bibfield  {journal} {\bibinfo
   {journal} {Phys. Rev. Lett.}\ }\textbf {\bibinfo {volume} {75}},\ \bibinfo
  {pages} {1260} (\bibinfo {year} {1995})},\ \Eprint
  {https://arxiv.org/abs/gr-qc/9504004} {arXiv:gr-qc/9504004} \BibitemShut
  {NoStop}%
\bibitem [{\citenamefont {Bekenstein}(1973)}]{PhysRevD.7.2333}%
  \BibitemOpen
  \bibfield  {author} {\bibinfo {author} {\bibfnamefont {J.~D.}\ \bibnamefont
  {Bekenstein}},\ }\bibfield  {title} {\bibinfo {title} {Black holes and
  entropy},\ }\href {https://doi.org/10.1103/PhysRevD.7.2333} {\bibfield
  {journal} {\bibinfo  {journal} {Phys. Rev. D}\ }\textbf {\bibinfo {volume}
  {7}},\ \bibinfo {pages} {2333} (\bibinfo {year} {1973})}\BibitemShut
  {NoStop}%
\bibitem [{\citenamefont {Bardeen}\ \emph {et~al.}(1973)\citenamefont
  {Bardeen}, \citenamefont {Carter},\ and\ \citenamefont
  {Hawking}}]{bardeen1973}%
  \BibitemOpen
  \bibfield  {author} {\bibinfo {author} {\bibfnamefont {J.~M.}\ \bibnamefont
  {Bardeen}}, \bibinfo {author} {\bibfnamefont {B.}~\bibnamefont {Carter}},\
  and\ \bibinfo {author} {\bibfnamefont {S.~W.}\ \bibnamefont {Hawking}},\
  }\bibfield  {title} {\bibinfo {title} {{The four laws of black hole
  mechanics}},\ }\href {https://doi.org/10.1007/BF01645742} {\bibfield
  {journal} {\bibinfo  {journal} {Commun. Math. Phys.}\ }\textbf {\bibinfo
  {volume} {31}},\ \bibinfo {pages} {161} (\bibinfo {year} {1973})}\BibitemShut
  {NoStop}%
\bibitem [{\citenamefont {Hawking}(1974)}]{hawking1974}%
  \BibitemOpen
  \bibfield  {author} {\bibinfo {author} {\bibfnamefont {S.~W.}\ \bibnamefont
  {Hawking}},\ }\bibfield  {title} {\bibinfo {title} {Black hole explosions?},\
  }\href {https://doi.org/10.1038/248030a0} {\bibfield  {journal} {\bibinfo
  {journal} {Nature}\ }\textbf {\bibinfo {volume} {248}},\ \bibinfo {pages}
  {30} (\bibinfo {year} {1974})}\BibitemShut {NoStop}%
\bibitem [{\citenamefont {{Hawking}}(1975)}]{1975CMaPh..43..199H}%
  \BibitemOpen
  \bibfield  {author} {\bibinfo {author} {\bibfnamefont {S.}~\bibnamefont
  {{Hawking}}},\ }\bibfield  {title} {\bibinfo {title} {{Particle creation by
  black holes}},\ }\href {https://doi.org/10.1007/BF02345020} {\bibfield
  {journal} {\bibinfo  {journal} {Commun. Math. Phys.}\ }\textbf {\bibinfo
  {volume} {43}},\ \bibinfo {pages} {199} (\bibinfo {year} {1975})}\BibitemShut
  {NoStop}%
\bibitem [{\citenamefont {Unruh}(1976)}]{PhysRevD.14.870}%
  \BibitemOpen
  \bibfield  {author} {\bibinfo {author} {\bibfnamefont {W.~G.}\ \bibnamefont
  {Unruh}},\ }\bibfield  {title} {\bibinfo {title} {Notes on black-hole
  evaporation},\ }\href {https://doi.org/10.1103/PhysRevD.14.870} {\bibfield
  {journal} {\bibinfo  {journal} {Phys. Rev. D}\ }\textbf {\bibinfo {volume}
  {14}},\ \bibinfo {pages} {870} (\bibinfo {year} {1976})}\BibitemShut
  {NoStop}%
\bibitem [{\citenamefont {Eling}\ \emph {et~al.}(2006)\citenamefont {Eling},
  \citenamefont {Guedens},\ and\ \citenamefont {Jacobson}}]{Eling:2006aw}%
  \BibitemOpen
  \bibfield  {author} {\bibinfo {author} {\bibfnamefont {C.}~\bibnamefont
  {Eling}}, \bibinfo {author} {\bibfnamefont {R.}~\bibnamefont {Guedens}},\
  and\ \bibinfo {author} {\bibfnamefont {T.}~\bibnamefont {Jacobson}},\
  }\bibfield  {title} {\bibinfo {title} {{Non-equilibrium thermodynamics of
  spacetime}},\ }\href {https://doi.org/10.1103/PhysRevLett.96.121301}
  {\bibfield  {journal} {\bibinfo  {journal} {Phys. Rev. Lett.}\ }\textbf
  {\bibinfo {volume} {96}},\ \bibinfo {pages} {121301} (\bibinfo {year}
  {2006})},\ \Eprint {https://arxiv.org/abs/gr-qc/0602001}
  {arXiv:gr-qc/0602001} \BibitemShut {NoStop}%
\bibitem [{\citenamefont {Lee}(2018)}]{lee2018introduction}%
  \BibitemOpen
  \bibfield  {author} {\bibinfo {author} {\bibfnamefont {J.~M.}\ \bibnamefont
  {Lee}},\ }\href {https://doi.org/10.1007/978-3-319-91755-9} {\emph {\bibinfo
  {title} {Introduction to Riemannian manifolds}}},\ Graduate Texts in
  Mathematics\ (\bibinfo  {publisher} {Springer},\ \bibinfo {year}
  {2018})\BibitemShut {NoStop}%
\bibitem [{\citenamefont {Misner}\ \emph {et~al.}(1973)\citenamefont {Misner},
  \citenamefont {Thorne},\ and\ \citenamefont
  {Wheeler}}]{thorne2000gravitation}%
  \BibitemOpen
  \bibfield  {author} {\bibinfo {author} {\bibfnamefont {C.~W.}\ \bibnamefont
  {Misner}}, \bibinfo {author} {\bibfnamefont {K.~S.}\ \bibnamefont {Thorne}},\
  and\ \bibinfo {author} {\bibfnamefont {J.~A.}\ \bibnamefont {Wheeler}},\
  }\href
  {https://press.princeton.edu/books/hardcover/9780691177793/gravitation}
  {\emph {\bibinfo {title} {Gravitation}}}\ (\bibinfo  {publisher} {W. H.
  Freeman and Company},\ \bibinfo {year} {1973})\BibitemShut {NoStop}%
\bibitem [{\citenamefont {Dey}\ \emph {et~al.}(2017)\citenamefont {Dey},
  \citenamefont {Liberati},\ and\ \citenamefont {Pranzetti}}]{Dey:2017fld}%
  \BibitemOpen
  \bibfield  {author} {\bibinfo {author} {\bibfnamefont {R.}~\bibnamefont
  {Dey}}, \bibinfo {author} {\bibfnamefont {S.}~\bibnamefont {Liberati}},\ and\
  \bibinfo {author} {\bibfnamefont {D.}~\bibnamefont {Pranzetti}},\ }\bibfield
  {title} {\bibinfo {title} {{Spacetime thermodynamics in the presence of
  torsion}},\ }\href {https://doi.org/10.1103/PhysRevD.96.124032} {\bibfield
  {journal} {\bibinfo  {journal} {Phys. Rev. D}\ }\textbf {\bibinfo {volume}
  {96}},\ \bibinfo {pages} {124032} (\bibinfo {year} {2017})},\ \Eprint
  {https://arxiv.org/abs/1709.04031} {arXiv:1709.04031 [gr-qc]} \BibitemShut
  {NoStop}%
\bibitem [{\citenamefont {De~Lorenzo}\ \emph {et~al.}(2018)\citenamefont
  {De~Lorenzo}, \citenamefont {De~Paoli},\ and\ \citenamefont
  {Speziale}}]{DeLorenzo:2018odq}%
  \BibitemOpen
  \bibfield  {author} {\bibinfo {author} {\bibfnamefont {T.}~\bibnamefont
  {De~Lorenzo}}, \bibinfo {author} {\bibfnamefont {E.}~\bibnamefont
  {De~Paoli}},\ and\ \bibinfo {author} {\bibfnamefont {S.}~\bibnamefont
  {Speziale}},\ }\bibfield  {title} {\bibinfo {title} {{Spacetime
  Thermodynamics with Contorsion}},\ }\href
  {https://doi.org/10.1103/PhysRevD.98.064053} {\bibfield  {journal} {\bibinfo
  {journal} {Phys. Rev. D}\ }\textbf {\bibinfo {volume} {98}},\ \bibinfo
  {pages} {064053} (\bibinfo {year} {2018})},\ \Eprint
  {https://arxiv.org/abs/1807.02041} {arXiv:1807.02041 [gr-qc]} \BibitemShut
  {NoStop}%
\bibitem [{\citenamefont {Cartan}(1923)}]{ASENS_1923_3_40__325_0}%
  \BibitemOpen
  \bibfield  {author} {\bibinfo {author} {\bibfnamefont {E.}~\bibnamefont
  {Cartan}},\ }\bibfield  {title} {\bibinfo {title} {Sur les vari\'et\'es \`a
  connexion affine et la th\'eorie de la relativit\'e g\'en\'eralis\'ee
  (premi\`ere partie)},\ }\href {https://doi.org/10.24033/asens.751} {\bibfield
   {journal} {\bibinfo  {journal} {Ann. Sci. de l'\'Ecole Norm.}\ }\textbf
  {\bibinfo {volume} {3e s{\'e}rie, 40}},\ \bibinfo {pages} {325} (\bibinfo
  {year} {1923})}\BibitemShut {NoStop}%
\bibitem [{\citenamefont {Blagojevi\'c}(2001)}]{blagojevic2001gravitation}%
  \BibitemOpen
  \bibfield  {author} {\bibinfo {author} {\bibfnamefont {M.}~\bibnamefont
  {Blagojevi\'c}},\ }\href {https://doi.org/10.1201/9781420034264} {\emph
  {\bibinfo {title} {Gravitation and Gauge Symmetries}}},\ Series in High
  Energy Physics, Cosmology and Gravitation\ (\bibinfo  {publisher} {CRC
  Press},\ \bibinfo {year} {2001})\BibitemShut {NoStop}%
\bibitem [{\citenamefont {Blagojevi{\'c}}\ and\ \citenamefont
  {Hehl}(2013)}]{Blagojevic:2013xpa}%
  \BibitemOpen
  \bibinfo {editor} {\bibfnamefont {M.}~\bibnamefont {Blagojevi{\'c}}}\ and\
  \bibinfo {editor} {\bibfnamefont {F.~W.}\ \bibnamefont {Hehl}},\ eds.,\ \href
  {https://doi.org/10.1142/p781} {\emph {\bibinfo {title} {{Gauge Theories of
  Gravitation}: {A Reader with Commentaries}}}}\ (\bibinfo  {publisher} {World
  Scientific},\ \bibinfo {year} {2013})\BibitemShut {NoStop}%
\bibitem [{\citenamefont {Hehl}\ \emph {et~al.}(1976)\citenamefont {Hehl},
  \citenamefont {Von Der~Heyde}, \citenamefont {Kerlick},\ and\ \citenamefont
  {Nester}}]{Hehl:1976kj}%
  \BibitemOpen
  \bibfield  {author} {\bibinfo {author} {\bibfnamefont {F.~W.}\ \bibnamefont
  {Hehl}}, \bibinfo {author} {\bibfnamefont {P.}~\bibnamefont {Von Der~Heyde}},
  \bibinfo {author} {\bibfnamefont {G.~D.}\ \bibnamefont {Kerlick}},\ and\
  \bibinfo {author} {\bibfnamefont {J.~M.}\ \bibnamefont {Nester}},\ }\bibfield
   {title} {\bibinfo {title} {{General Relativity with Spin and Torsion:
  Foundations and Prospects}},\ }\href
  {https://doi.org/10.1103/RevModPhys.48.393} {\bibfield  {journal} {\bibinfo
  {journal} {Rev. Mod. Phys.}\ }\textbf {\bibinfo {volume} {48}},\ \bibinfo
  {pages} {393} (\bibinfo {year} {1976})}\BibitemShut {NoStop}%
\bibitem [{\citenamefont {Hehl}(1979)}]{Hehl1979}%
  \BibitemOpen
  \bibfield  {author} {\bibinfo {author} {\bibfnamefont {F.~W.}\ \bibnamefont
  {Hehl}},\ }\bibfield  {title} {\bibinfo {title} {Four lectures on
  \text{Poincaré} gauge field theory},\ }\href
  {https://doi.org/10.1007/978-1-4613-3123-0_2} {\bibfield  {journal} {\bibinfo
   {journal} {Nato Science Series B}\ }\textbf {\bibinfo {volume} {58}},\
  \bibinfo {pages} {5} (\bibinfo {year} {1979})}\BibitemShut {NoStop}%
\bibitem [{\citenamefont {Smolin}(2014)}]{Smolin:2012ys}%
  \BibitemOpen
  \bibfield  {author} {\bibinfo {author} {\bibfnamefont {L.}~\bibnamefont
  {Smolin}},\ }\bibfield  {title} {\bibinfo {title} {{General relativity as the
  equation of state of spin foam}},\ }\href
  {https://doi.org/10.1088/0264-9381/31/19/195007} {\bibfield  {journal}
  {\bibinfo  {journal} {Class. Quant. Grav.}\ }\textbf {\bibinfo {volume}
  {31}},\ \bibinfo {pages} {195007} (\bibinfo {year} {2014})},\ \Eprint
  {https://arxiv.org/abs/1205.5529} {arXiv:1205.5529 [gr-qc]} \BibitemShut
  {NoStop}%
\bibitem [{\citenamefont {Chirco}\ \emph {et~al.}(2014)\citenamefont {Chirco},
  \citenamefont {Haggard}, \citenamefont {Riello},\ and\ \citenamefont
  {Rovelli}}]{Chirco:2014saa}%
  \BibitemOpen
  \bibfield  {author} {\bibinfo {author} {\bibfnamefont {G.}~\bibnamefont
  {Chirco}}, \bibinfo {author} {\bibfnamefont {H.~M.}\ \bibnamefont {Haggard}},
  \bibinfo {author} {\bibfnamefont {A.}~\bibnamefont {Riello}},\ and\ \bibinfo
  {author} {\bibfnamefont {C.}~\bibnamefont {Rovelli}},\ }\bibfield  {title}
  {\bibinfo {title} {{Spacetime thermodynamics without hidden degrees of
  freedom}},\ }\href {https://doi.org/10.1103/PhysRevD.90.044044} {\bibfield
  {journal} {\bibinfo  {journal} {Phys. Rev. D}\ }\textbf {\bibinfo {volume}
  {90}},\ \bibinfo {pages} {044044} (\bibinfo {year} {2014})},\ \Eprint
  {https://arxiv.org/abs/1401.5262} {arXiv:1401.5262 [gr-qc]} \BibitemShut
  {NoStop}%
\bibitem [{\citenamefont {Gielen}\ \emph {et~al.}(2013)\citenamefont {Gielen},
  \citenamefont {Oriti},\ and\ \citenamefont {Sindoni}}]{Gielen:2013kla}%
  \BibitemOpen
  \bibfield  {author} {\bibinfo {author} {\bibfnamefont {S.}~\bibnamefont
  {Gielen}}, \bibinfo {author} {\bibfnamefont {D.}~\bibnamefont {Oriti}},\ and\
  \bibinfo {author} {\bibfnamefont {L.}~\bibnamefont {Sindoni}},\ }\bibfield
  {title} {\bibinfo {title} {{Cosmology from Group Field Theory Formalism for
  Quantum Gravity}},\ }\href {https://doi.org/10.1103/PhysRevLett.111.031301}
  {\bibfield  {journal} {\bibinfo  {journal} {Phys. Rev. Lett.}\ }\textbf
  {\bibinfo {volume} {111}},\ \bibinfo {pages} {031301} (\bibinfo {year}
  {2013})},\ \Eprint {https://arxiv.org/abs/1303.3576} {arXiv:1303.3576
  [gr-qc]} \BibitemShut {NoStop}%
\bibitem [{\citenamefont {Oriti}\ \emph {et~al.}(2015)\citenamefont {Oriti},
  \citenamefont {Pranzetti}, \citenamefont {Ryan},\ and\ \citenamefont
  {Sindoni}}]{Oriti:2015qva}%
  \BibitemOpen
  \bibfield  {author} {\bibinfo {author} {\bibfnamefont {D.}~\bibnamefont
  {Oriti}}, \bibinfo {author} {\bibfnamefont {D.}~\bibnamefont {Pranzetti}},
  \bibinfo {author} {\bibfnamefont {J.~P.}\ \bibnamefont {Ryan}},\ and\
  \bibinfo {author} {\bibfnamefont {L.}~\bibnamefont {Sindoni}},\ }\bibfield
  {title} {\bibinfo {title} {{Generalized quantum gravity condensates for
  homogeneous geometries and cosmology}},\ }\href
  {https://doi.org/10.1088/0264-9381/32/23/235016} {\bibfield  {journal}
  {\bibinfo  {journal} {Class. Quant. Grav.}\ }\textbf {\bibinfo {volume}
  {32}},\ \bibinfo {pages} {235016} (\bibinfo {year} {2015})},\ \Eprint
  {https://arxiv.org/abs/1501.00936} {arXiv:1501.00936 [gr-qc]} \BibitemShut
  {NoStop}%
\bibitem [{\citenamefont {Oriti}\ \emph
  {et~al.}(2016{\natexlab{a}})\citenamefont {Oriti}, \citenamefont
  {Pranzetti},\ and\ \citenamefont {Sindoni}}]{Oriti:2015rwa}%
  \BibitemOpen
  \bibfield  {author} {\bibinfo {author} {\bibfnamefont {D.}~\bibnamefont
  {Oriti}}, \bibinfo {author} {\bibfnamefont {D.}~\bibnamefont {Pranzetti}},\
  and\ \bibinfo {author} {\bibfnamefont {L.}~\bibnamefont {Sindoni}},\
  }\bibfield  {title} {\bibinfo {title} {{Horizon entropy from quantum gravity
  condensates}},\ }\href {https://doi.org/10.1103/PhysRevLett.116.211301}
  {\bibfield  {journal} {\bibinfo  {journal} {Phys. Rev. Lett.}\ }\textbf
  {\bibinfo {volume} {116}},\ \bibinfo {pages} {211301} (\bibinfo {year}
  {2016}{\natexlab{a}})},\ \Eprint {https://arxiv.org/abs/1510.06991}
  {arXiv:1510.06991 [gr-qc]} \BibitemShut {NoStop}%
\bibitem [{\citenamefont {Oriti}\ \emph
  {et~al.}(2016{\natexlab{b}})\citenamefont {Oriti}, \citenamefont {Sindoni},\
  and\ \citenamefont {Wilson-Ewing}}]{Oriti:2016qtz}%
  \BibitemOpen
  \bibfield  {author} {\bibinfo {author} {\bibfnamefont {D.}~\bibnamefont
  {Oriti}}, \bibinfo {author} {\bibfnamefont {L.}~\bibnamefont {Sindoni}},\
  and\ \bibinfo {author} {\bibfnamefont {E.}~\bibnamefont {Wilson-Ewing}},\
  }\bibfield  {title} {\bibinfo {title} {{Emergent Friedmann dynamics with a
  quantum bounce from quantum gravity condensates}},\ }\href
  {https://doi.org/10.1088/0264-9381/33/22/224001} {\bibfield  {journal}
  {\bibinfo  {journal} {Class. Quant. Grav.}\ }\textbf {\bibinfo {volume}
  {33}},\ \bibinfo {pages} {224001} (\bibinfo {year} {2016}{\natexlab{b}})},\
  \Eprint {https://arxiv.org/abs/1602.05881} {arXiv:1602.05881 [gr-qc]}
  \BibitemShut {NoStop}%
\bibitem [{\citenamefont {Dittrich}\ and\ \citenamefont
  {Steinhaus}(2014)}]{Dittrich:2013xwa}%
  \BibitemOpen
  \bibfield  {author} {\bibinfo {author} {\bibfnamefont {B.}~\bibnamefont
  {Dittrich}}\ and\ \bibinfo {author} {\bibfnamefont {S.}~\bibnamefont
  {Steinhaus}},\ }\bibfield  {title} {\bibinfo {title} {{Time evolution as
  refining, coarse graining and entangling}},\ }\href
  {https://doi.org/10.1088/1367-2630/16/12/123041} {\bibfield  {journal}
  {\bibinfo  {journal} {New J. Phys.}\ }\textbf {\bibinfo {volume} {16}},\
  \bibinfo {pages} {123041} (\bibinfo {year} {2014})},\ \Eprint
  {https://arxiv.org/abs/1311.7565} {arXiv:1311.7565 [gr-qc]} \BibitemShut
  {NoStop}%
\bibitem [{\citenamefont {Dittrich}\ \emph
  {et~al.}(2016{\natexlab{a}})\citenamefont {Dittrich}, \citenamefont
  {Mizera},\ and\ \citenamefont {Steinhaus}}]{Dittrich:2014mxa}%
  \BibitemOpen
  \bibfield  {author} {\bibinfo {author} {\bibfnamefont {B.}~\bibnamefont
  {Dittrich}}, \bibinfo {author} {\bibfnamefont {S.}~\bibnamefont {Mizera}},\
  and\ \bibinfo {author} {\bibfnamefont {S.}~\bibnamefont {Steinhaus}},\
  }\bibfield  {title} {\bibinfo {title} {{Decorated tensor network
  renormalization for lattice gauge theories and spin foam models}},\ }\href
  {https://doi.org/10.1088/1367-2630/18/5/053009} {\bibfield  {journal}
  {\bibinfo  {journal} {New J. Phys.}\ }\textbf {\bibinfo {volume} {18}},\
  \bibinfo {pages} {053009} (\bibinfo {year} {2016}{\natexlab{a}})},\ \Eprint
  {https://arxiv.org/abs/1409.2407} {arXiv:1409.2407 [gr-qc]} \BibitemShut
  {NoStop}%
\bibitem [{\citenamefont {Dittrich}\ \emph
  {et~al.}(2016{\natexlab{b}})\citenamefont {Dittrich}, \citenamefont
  {Schnetter}, \citenamefont {Seth},\ and\ \citenamefont
  {Steinhaus}}]{Dittrich:2016tys}%
  \BibitemOpen
  \bibfield  {author} {\bibinfo {author} {\bibfnamefont {B.}~\bibnamefont
  {Dittrich}}, \bibinfo {author} {\bibfnamefont {E.}~\bibnamefont {Schnetter}},
  \bibinfo {author} {\bibfnamefont {C.~J.}\ \bibnamefont {Seth}},\ and\
  \bibinfo {author} {\bibfnamefont {S.}~\bibnamefont {Steinhaus}},\ }\bibfield
  {title} {\bibinfo {title} {{Coarse graining flow of spin foam
  intertwiners}},\ }\href {https://doi.org/10.1103/PhysRevD.94.124050}
  {\bibfield  {journal} {\bibinfo  {journal} {Phys. Rev. D}\ }\textbf {\bibinfo
  {volume} {94}},\ \bibinfo {pages} {124050} (\bibinfo {year}
  {2016}{\natexlab{b}})},\ \Eprint {https://arxiv.org/abs/1609.02429}
  {arXiv:1609.02429 [gr-qc]} \BibitemShut {NoStop}%
\bibitem [{\citenamefont {Dey}\ \emph {et~al.}(2016)\citenamefont {Dey},
  \citenamefont {Liberati},\ and\ \citenamefont {Mohd}}]{Dey:2016zka}%
  \BibitemOpen
  \bibfield  {author} {\bibinfo {author} {\bibfnamefont {R.}~\bibnamefont
  {Dey}}, \bibinfo {author} {\bibfnamefont {S.}~\bibnamefont {Liberati}},\ and\
  \bibinfo {author} {\bibfnamefont {A.}~\bibnamefont {Mohd}},\ }\bibfield
  {title} {\bibinfo {title} {{Higher derivative gravity: field equation as the
  equation of state}},\ }\href {https://doi.org/10.1103/PhysRevD.94.044013}
  {\bibfield  {journal} {\bibinfo  {journal} {Phys. Rev. D}\ }\textbf {\bibinfo
  {volume} {94}},\ \bibinfo {pages} {044013} (\bibinfo {year} {2016})},\
  \Eprint {https://arxiv.org/abs/1605.04789} {arXiv:1605.04789 [gr-qc]}
  \BibitemShut {NoStop}%
\bibitem [{\citenamefont {Guedens}\ \emph {et~al.}(2012)\citenamefont
  {Guedens}, \citenamefont {Jacobson},\ and\ \citenamefont
  {Sarkar}}]{Guedens:2011dy}%
  \BibitemOpen
  \bibfield  {author} {\bibinfo {author} {\bibfnamefont {R.}~\bibnamefont
  {Guedens}}, \bibinfo {author} {\bibfnamefont {T.}~\bibnamefont {Jacobson}},\
  and\ \bibinfo {author} {\bibfnamefont {S.}~\bibnamefont {Sarkar}},\
  }\bibfield  {title} {\bibinfo {title} {{Horizon entropy and higher curvature
  equations of state}},\ }\href {https://doi.org/10.1103/PhysRevD.85.064017}
  {\bibfield  {journal} {\bibinfo  {journal} {Phys. Rev. D}\ }\textbf {\bibinfo
  {volume} {85}},\ \bibinfo {pages} {064017} (\bibinfo {year} {2012})},\
  \Eprint {https://arxiv.org/abs/1112.6215} {arXiv:1112.6215 [gr-qc]}
  \BibitemShut {NoStop}%
\bibitem [{\citenamefont {Hehl}\ \emph {et~al.}(1995)\citenamefont {Hehl},
  \citenamefont {McCrea}, \citenamefont {Mielke},\ and\ \citenamefont
  {Ne'eman}}]{Hehl:1994ue}%
  \BibitemOpen
  \bibfield  {author} {\bibinfo {author} {\bibfnamefont {F.~W.}\ \bibnamefont
  {Hehl}}, \bibinfo {author} {\bibfnamefont {J.~D.}\ \bibnamefont {McCrea}},
  \bibinfo {author} {\bibfnamefont {E.~W.}\ \bibnamefont {Mielke}},\ and\
  \bibinfo {author} {\bibfnamefont {Y.}~\bibnamefont {Ne'eman}},\ }\bibfield
  {title} {\bibinfo {title} {{Metric affine gauge theory of gravity: Field
  equations, Noether identities, world spinors, and breaking of dilation
  invariance}},\ }\href {https://doi.org/10.1016/0370-1573(94)00111-F}
  {\bibfield  {journal} {\bibinfo  {journal} {Phys. Rept.}\ }\textbf {\bibinfo
  {volume} {258}},\ \bibinfo {pages} {1} (\bibinfo {year} {1995})},\ \Eprint
  {https://arxiv.org/abs/gr-qc/9402012} {arXiv:gr-qc/9402012} \BibitemShut
  {NoStop}%
\bibitem [{\citenamefont {Jiménez-Cano}(2022)}]{JimenezCano:2021rlu}%
  \BibitemOpen
  \bibfield  {author} {\bibinfo {author} {\bibfnamefont {A.}~\bibnamefont
  {Jiménez-Cano}},\ }\bibfield  {title} {\bibinfo {title} {{Metric-Affine
  Gauge theories of gravity: Foundations and new insights}},\ }\href@noop {}
  {\bibfield  {journal} {\bibinfo  {journal} {Granada U., Theor. Phys.
  Astrophys.}\ } (\bibinfo {year} {2022})},\ \Eprint
  {https://arxiv.org/abs/2201.12847} {arXiv:2201.12847 [gr-qc]} \BibitemShut
  {NoStop}%
\bibitem [{\citenamefont {Andrei}\ \emph {et~al.}(2025)\citenamefont {Andrei},
  \citenamefont {Iosifidis}, \citenamefont {J{\"a}rv},\ and\ \citenamefont
  {Saal}}]{Andrei:2024vvy}%
  \BibitemOpen
  \bibfield  {author} {\bibinfo {author} {\bibfnamefont {I.}~\bibnamefont
  {Andrei}}, \bibinfo {author} {\bibfnamefont {D.}~\bibnamefont {Iosifidis}},
  \bibinfo {author} {\bibfnamefont {L.}~\bibnamefont {J{\"a}rv}},\ and\
  \bibinfo {author} {\bibfnamefont {M.}~\bibnamefont {Saal}},\ }\bibfield
  {title} {\bibinfo {title} {{Friedmann cosmology with hyperfluids}},\ }\href
  {https://doi.org/10.1103/PhysRevD.111.064063} {\bibfield  {journal} {\bibinfo
   {journal} {Phys. Rev. D}\ }\textbf {\bibinfo {volume} {111}},\ \bibinfo
  {pages} {064063} (\bibinfo {year} {2025})},\ \Eprint
  {https://arxiv.org/abs/2411.19127} {arXiv:2411.19127 [gr-qc]} \BibitemShut
  {NoStop}%
\bibitem [{\citenamefont {Falk}(1981)}]{falk1981}%
  \BibitemOpen
  \bibfield  {author} {\bibinfo {author} {\bibfnamefont {F.}~\bibnamefont
  {Falk}},\ }\bibfield  {title} {\bibinfo {title} {{Theory of elasticity of
  coherent inclusions by means of non-metric geometry}},\ }\href
  {https://doi.org/https://doi.org/10.1007/BF00058079} {\bibfield  {journal}
  {\bibinfo  {journal} {J. Elast.}\ }\textbf {\bibinfo {volume} {11}},\
  \bibinfo {pages} {359} (\bibinfo {year} {1981})}\BibitemShut {NoStop}%
\bibitem [{\citenamefont {Kupferman}\ and\ \citenamefont
  {Maor}(2015)}]{Kupferman2015}%
  \BibitemOpen
  \bibfield  {author} {\bibinfo {author} {\bibfnamefont {R.}~\bibnamefont
  {Kupferman}}\ and\ \bibinfo {author} {\bibfnamefont {C.}~\bibnamefont
  {Maor}},\ }\bibfield  {title} {\bibinfo {title} {The emergence of torsion in
  the continuum limit of distributed edge-dislocations},\ }\href
  {https://doi.org/10.3934/jgm.2015.7.361} {\bibfield  {journal} {\bibinfo
  {journal} {J. Geom. Mech.}\ }\textbf {\bibinfo {volume} {7}},\ \bibinfo
  {pages} {361} (\bibinfo {year} {2015})},\ \Eprint
  {https://arxiv.org/abs/1410.2906} {arXiv:1410.2906 [math.DG]} \BibitemShut
  {NoStop}%
\bibitem [{\citenamefont {Kupferman}\ \emph {et~al.}(2017)\citenamefont
  {Kupferman}, \citenamefont {Maor},\ and\ \citenamefont
  {Rosenthal}}]{Kupferman2017}%
  \BibitemOpen
  \bibfield  {author} {\bibinfo {author} {\bibfnamefont {R.}~\bibnamefont
  {Kupferman}}, \bibinfo {author} {\bibfnamefont {C.}~\bibnamefont {Maor}},\
  and\ \bibinfo {author} {\bibfnamefont {R.}~\bibnamefont {Rosenthal}},\
  }\bibfield  {title} {\bibinfo {title} {Non-metricity in the continuum limit
  of randomly-distributed point defects},\ }\href
  {https://doi.org/10.1007/s11856-017-1620-x} {\bibfield  {journal} {\bibinfo
  {journal} {Isr. J. Math.}\ }\textbf {\bibinfo {volume} {223}},\ \bibinfo
  {pages} {75} (\bibinfo {year} {2017})},\ \Eprint
  {https://arxiv.org/abs/1508.02003} {arXiv:1508.02003 [math.PR]} \BibitemShut
  {NoStop}%
\bibitem [{\citenamefont {Lanczos}(1932)}]{Lanczos:1932fxw}%
  \BibitemOpen
  \bibfield  {author} {\bibinfo {author} {\bibfnamefont {C.}~\bibnamefont
  {Lanczos}},\ }\bibfield  {title} {\bibinfo {title} {{Elektromagnetismus als
  nat{\"u}rliche Eigenschaft der Riemannschen Geometrie}},\ }\href
  {https://doi.org/10.1007/BF01351210} {\bibfield  {journal} {\bibinfo
  {journal} {Z. Phys.}\ }\textbf {\bibinfo {volume} {73}},\ \bibinfo {pages}
  {147} (\bibinfo {year} {1932})}\BibitemShut {NoStop}%
\bibitem [{\citenamefont {Lanczos}(1938)}]{Lanczos:1938sf}%
  \BibitemOpen
  \bibfield  {author} {\bibinfo {author} {\bibfnamefont {C.}~\bibnamefont
  {Lanczos}},\ }\bibfield  {title} {\bibinfo {title} {{A Remarkable property of
  the Riemann-Christoffel tensor in four dimensions}},\ }\href
  {https://doi.org/10.2307/1968467} {\bibfield  {journal} {\bibinfo  {journal}
  {Annals Math.}\ }\textbf {\bibinfo {volume} {39}},\ \bibinfo {pages} {842}
  (\bibinfo {year} {1938})}\BibitemShut {NoStop}%
\bibitem [{\citenamefont {Lovelock}(1969)}]{Lovelock:1969vyr}%
  \BibitemOpen
  \bibfield  {author} {\bibinfo {author} {\bibfnamefont {D.}~\bibnamefont
  {Lovelock}},\ }\bibfield  {title} {\bibinfo {title} {{The uniqueness of the
  Einstein field equations in a four-dimensional space}},\ }\href
  {https://doi.org/10.1007/BF00248156} {\bibfield  {journal} {\bibinfo
  {journal} {Arch. Ration. Mech. Anal.}\ }\textbf {\bibinfo {volume} {33}},\
  \bibinfo {pages} {54} (\bibinfo {year} {1969})}\BibitemShut {NoStop}%
\bibitem [{\citenamefont {Lovelock}(1970)}]{Lovelock:1970zsf}%
  \BibitemOpen
  \bibfield  {author} {\bibinfo {author} {\bibfnamefont {D.}~\bibnamefont
  {Lovelock}},\ }\bibfield  {title} {\bibinfo {title} {{Divergence-free
  tensorial concomitants}},\ }\href {https://doi.org/10.1007/BF01817753}
  {\bibfield  {journal} {\bibinfo  {journal} {Aequat. Math.}\ }\textbf
  {\bibinfo {volume} {4}},\ \bibinfo {pages} {127} (\bibinfo {year}
  {1970})}\BibitemShut {NoStop}%
\bibitem [{\citenamefont {Lovelock}(1971)}]{Lovelock:1971yv}%
  \BibitemOpen
  \bibfield  {author} {\bibinfo {author} {\bibfnamefont {D.}~\bibnamefont
  {Lovelock}},\ }\bibfield  {title} {\bibinfo {title} {{The Einstein tensor and
  its generalizations}},\ }\href {https://doi.org/10.1063/1.1665613} {\bibfield
   {journal} {\bibinfo  {journal} {J. Math. Phys.}\ }\textbf {\bibinfo {volume}
  {12}},\ \bibinfo {pages} {498} (\bibinfo {year} {1971})}\BibitemShut
  {NoStop}%
\bibitem [{\citenamefont {Lovelock}(1972)}]{Lovelock:1972vz}%
  \BibitemOpen
  \bibfield  {author} {\bibinfo {author} {\bibfnamefont {D.}~\bibnamefont
  {Lovelock}},\ }\bibfield  {title} {\bibinfo {title} {{The four-dimensionality
  of space and the Einstein tensor}},\ }\href
  {https://doi.org/10.1063/1.1666069} {\bibfield  {journal} {\bibinfo
  {journal} {J. Math. Phys.}\ }\textbf {\bibinfo {volume} {13}},\ \bibinfo
  {pages} {874} (\bibinfo {year} {1972})}\BibitemShut {NoStop}%
\bibitem [{\citenamefont {Pardo}\ \emph {et~al.}(2018)\citenamefont {Pardo},
  \citenamefont {Fishbach}, \citenamefont {Holz},\ and\ \citenamefont
  {Spergel}}]{Pardo:2018ipy}%
  \BibitemOpen
  \bibfield  {author} {\bibinfo {author} {\bibfnamefont {K.}~\bibnamefont
  {Pardo}}, \bibinfo {author} {\bibfnamefont {M.}~\bibnamefont {Fishbach}},
  \bibinfo {author} {\bibfnamefont {D.~E.}\ \bibnamefont {Holz}},\ and\
  \bibinfo {author} {\bibfnamefont {D.~N.}\ \bibnamefont {Spergel}},\
  }\bibfield  {title} {\bibinfo {title} {{Limits on the number of spacetime
  dimensions from GW170817}},\ }\href
  {https://doi.org/10.1088/1475-7516/2018/07/048} {\bibfield  {journal}
  {\bibinfo  {journal} {JCAP}\ }\textbf {\bibinfo {volume} {07}}\bibfield
  {number} {\bibinfo  {number} { (2018)},\ \bibinfo {pages} {048}},\ }\Eprint
  {https://arxiv.org/abs/1801.08160} {arXiv:1801.08160 [gr-qc]} \BibitemShut
  {NoStop}%
\bibitem [{\citenamefont {Schouten}(1954)}]{schouten2013ricci}%
  \BibitemOpen
  \bibfield  {author} {\bibinfo {author} {\bibfnamefont {J.}~\bibnamefont
  {Schouten}},\ }\href {https://doi.org/10.1007/978-3-662-12927-2} {\emph
  {\bibinfo {title} {Ricci-Calculus: An Introduction to Tensor Analysis and Its
  Geometrical Applications}}},\ Grundlehren der mathematischen Wissenschaften\
  (\bibinfo  {publisher} {Springer},\ \bibinfo {year} {1954})\BibitemShut
  {NoStop}%
\bibitem [{\citenamefont {Iosifidis}\ \emph {et~al.}(2018)\citenamefont
  {Iosifidis}, \citenamefont {Tsagas},\ and\ \citenamefont
  {Petkou}}]{Iosifidis:2018diy}%
  \BibitemOpen
  \bibfield  {author} {\bibinfo {author} {\bibfnamefont {D.}~\bibnamefont
  {Iosifidis}}, \bibinfo {author} {\bibfnamefont {C.~G.}\ \bibnamefont
  {Tsagas}},\ and\ \bibinfo {author} {\bibfnamefont {A.~C.}\ \bibnamefont
  {Petkou}},\ }\bibfield  {title} {\bibinfo {title} {{Raychaudhuri equation in
  spacetimes with torsion and nonmetricity}},\ }\href
  {https://doi.org/10.1103/PhysRevD.98.104037} {\bibfield  {journal} {\bibinfo
  {journal} {Phys. Rev. D}\ }\textbf {\bibinfo {volume} {98}},\ \bibinfo
  {pages} {104037} (\bibinfo {year} {2018})},\ \Eprint
  {https://arxiv.org/abs/1809.04992} {arXiv:1809.04992 [gr-qc]} \BibitemShut
  {NoStop}%
\bibitem [{\citenamefont {'t~Hooft}(1993)}]{tHooft:1993dmi}%
  \BibitemOpen
  \bibfield  {author} {\bibinfo {author} {\bibfnamefont {G.}~\bibnamefont
  {'t~Hooft}},\ }\bibfield  {title} {\bibinfo {title} {{Dimensional reduction
  in quantum gravity}},\ }\href@noop {} {\bibfield  {journal} {\bibinfo
  {journal} {Conf. Proc. C}\ }\textbf {\bibinfo {volume} {930308}},\ \bibinfo
  {pages} {284} (\bibinfo {year} {1993})},\ \Eprint
  {https://arxiv.org/abs/gr-qc/9310026} {arXiv:gr-qc/9310026} \BibitemShut
  {NoStop}%
\bibitem [{\citenamefont {Susskind}(1995)}]{Susskind:1994vu}%
  \BibitemOpen
  \bibfield  {author} {\bibinfo {author} {\bibfnamefont {L.}~\bibnamefont
  {Susskind}},\ }\bibfield  {title} {\bibinfo {title} {{The World as a
  hologram}},\ }\href {https://doi.org/10.1063/1.531249} {\bibfield  {journal}
  {\bibinfo  {journal} {J. Math. Phys.}\ }\textbf {\bibinfo {volume} {36}},\
  \bibinfo {pages} {6377} (\bibinfo {year} {1995})},\ \Eprint
  {https://arxiv.org/abs/hep-th/9409089} {arXiv:hep-th/9409089} \BibitemShut
  {NoStop}%
\bibitem [{\citenamefont {Bombelli}\ \emph {et~al.}(1986)\citenamefont
  {Bombelli}, \citenamefont {Koul}, \citenamefont {Lee},\ and\ \citenamefont
  {Sorkin}}]{PhysRevD.34.373}%
  \BibitemOpen
  \bibfield  {author} {\bibinfo {author} {\bibfnamefont {L.}~\bibnamefont
  {Bombelli}}, \bibinfo {author} {\bibfnamefont {R.~K.}\ \bibnamefont {Koul}},
  \bibinfo {author} {\bibfnamefont {J.}~\bibnamefont {Lee}},\ and\ \bibinfo
  {author} {\bibfnamefont {R.~D.}\ \bibnamefont {Sorkin}},\ }\bibfield  {title}
  {\bibinfo {title} {Quantum source of entropy for black holes},\ }\href
  {https://doi.org/10.1103/PhysRevD.34.373} {\bibfield  {journal} {\bibinfo
  {journal} {Phys. Rev. D}\ }\textbf {\bibinfo {volume} {34}},\ \bibinfo
  {pages} {373} (\bibinfo {year} {1986})}\BibitemShut {NoStop}%
\bibitem [{\citenamefont {Blaschke}\ \emph {et~al.}(2016)\citenamefont
  {Blaschke}, \citenamefont {Gieres}, \citenamefont {Reboud},\ and\
  \citenamefont {Schweda}}]{Blaschke:2016ohs}%
  \BibitemOpen
  \bibfield  {author} {\bibinfo {author} {\bibfnamefont {D.~N.}\ \bibnamefont
  {Blaschke}}, \bibinfo {author} {\bibfnamefont {F.}~\bibnamefont {Gieres}},
  \bibinfo {author} {\bibfnamefont {M.}~\bibnamefont {Reboud}},\ and\ \bibinfo
  {author} {\bibfnamefont {M.}~\bibnamefont {Schweda}},\ }\bibfield  {title}
  {\bibinfo {title} {{The energy{\textendash}momentum tensor(s) in classical
  gauge theories}},\ }\href {https://doi.org/10.1016/j.nuclphysb.2016.07.001}
  {\bibfield  {journal} {\bibinfo  {journal} {Nucl. Phys. B}\ }\textbf
  {\bibinfo {volume} {912}},\ \bibinfo {pages} {192} (\bibinfo {year}
  {2016})},\ \Eprint {https://arxiv.org/abs/1605.01121} {arXiv:1605.01121
  [hep-th]} \BibitemShut {NoStop}%
\bibitem [{\citenamefont {Baker}\ \emph {et~al.}(2021)\citenamefont {Baker},
  \citenamefont {Kiriushcheva},\ and\ \citenamefont {Kuzmin}}]{Baker:2020eqs}%
  \BibitemOpen
  \bibfield  {author} {\bibinfo {author} {\bibfnamefont {M.~R.}\ \bibnamefont
  {Baker}}, \bibinfo {author} {\bibfnamefont {N.}~\bibnamefont
  {Kiriushcheva}},\ and\ \bibinfo {author} {\bibfnamefont {S.}~\bibnamefont
  {Kuzmin}},\ }\bibfield  {title} {\bibinfo {title} {{Noether and Hilbert
  (metric) energy-momentum tensors are not, in general, equivalent}},\ }\href
  {https://doi.org/10.1016/j.nuclphysb.2020.115240} {\bibfield  {journal}
  {\bibinfo  {journal} {Nucl. Phys. B}\ }\textbf {\bibinfo {volume} {962}},\
  \bibinfo {pages} {115240} (\bibinfo {year} {2021})},\ \Eprint
  {https://arxiv.org/abs/2011.10611} {arXiv:2011.10611 [math-ph]} \BibitemShut
  {NoStop}%
\bibitem [{\citenamefont {Mardones}\ and\ \citenamefont
  {Zanelli}(1991)}]{AMar1991}%
  \BibitemOpen
  \bibfield  {author} {\bibinfo {author} {\bibfnamefont {A.}~\bibnamefont
  {Mardones}}\ and\ \bibinfo {author} {\bibfnamefont {J.}~\bibnamefont
  {Zanelli}},\ }\bibfield  {title} {\bibinfo {title} {\text{Lovelock-Cartan
  theory of gravity}},\ }\href {https://doi.org/10.1088/0264-9381/8/8/018}
  {\bibfield  {journal} {\bibinfo  {journal} {Class. Quant. Grav.}\ }\textbf
  {\bibinfo {volume} {8}},\ \bibinfo {pages} {1545} (\bibinfo {year}
  {1991})}\BibitemShut {NoStop}%
\bibitem [{\citenamefont {Iosifidis}\ and\ \citenamefont
  {Hehl}(2024)}]{Iosifidis:2023eom}%
  \BibitemOpen
  \bibfield  {author} {\bibinfo {author} {\bibfnamefont {D.}~\bibnamefont
  {Iosifidis}}\ and\ \bibinfo {author} {\bibfnamefont {F.~W.}\ \bibnamefont
  {Hehl}},\ }\bibfield  {title} {\bibinfo {title} {{Motion of test particles in
  spacetimes with torsion and nonmetricity}},\ }\href
  {https://doi.org/10.1016/j.physletb.2024.138498} {\bibfield  {journal}
  {\bibinfo  {journal} {Phys. Lett. B}\ }\textbf {\bibinfo {volume} {850}},\
  \bibinfo {pages} {138498} (\bibinfo {year} {2024})},\ \Eprint
  {https://arxiv.org/abs/2310.15595} {arXiv:2310.15595 [gr-qc]} \BibitemShut
  {NoStop}%
\bibitem [{\citenamefont {Iosifidis}(2022)}]{Iosifidis:2021bad}%
  \BibitemOpen
  \bibfield  {author} {\bibinfo {author} {\bibfnamefont {D.}~\bibnamefont
  {Iosifidis}},\ }\bibfield  {title} {\bibinfo {title} {{The full quadratic
  metric-affine gravity (including parity odd terms): exact solutions for the
  affine-connection}},\ }\href {https://doi.org/10.1088/1361-6382/ac6058}
  {\bibfield  {journal} {\bibinfo  {journal} {Class. Quant. Grav.}\ }\textbf
  {\bibinfo {volume} {39}},\ \bibinfo {pages} {095002} (\bibinfo {year}
  {2022})},\ \Eprint {https://arxiv.org/abs/2112.09154} {arXiv:2112.09154
  [gr-qc]} \BibitemShut {NoStop}%
\bibitem [{\citenamefont {Iosifidis}(2021)}]{Iosifidis:2021nra}%
  \BibitemOpen
  \bibfield  {author} {\bibinfo {author} {\bibfnamefont {D.}~\bibnamefont
  {Iosifidis}},\ }\bibfield  {title} {\bibinfo {title} {{The Perfect Hyperfluid
  of Metric-Affine Gravity: The Foundation}},\ }\href
  {https://doi.org/10.1088/1475-7516/2021/04/072} {\bibfield  {journal}
  {\bibinfo  {journal} {JCAP}\ }\textbf {\bibinfo {volume} {04}}\bibfield
  {number} {\bibinfo  {number} { (2021)},\ \bibinfo {pages} {072}},\ }\Eprint
  {https://arxiv.org/abs/2101.07289} {arXiv:2101.07289 [gr-qc]} \BibitemShut
  {NoStop}%
\bibitem [{\citenamefont {Iosifidis}(2020)}]{Iosifidis:2020gth}%
  \BibitemOpen
  \bibfield  {author} {\bibinfo {author} {\bibfnamefont {D.}~\bibnamefont
  {Iosifidis}},\ }\bibfield  {title} {\bibinfo {title} {{Cosmological
  Hyperfluids, Torsion and Non-metricity}},\ }\href
  {https://doi.org/10.1140/epjc/s10052-020-08634-z} {\bibfield  {journal}
  {\bibinfo  {journal} {Eur. Phys. J. C}\ }\textbf {\bibinfo {volume} {80}},\
  \bibinfo {pages} {1042} (\bibinfo {year} {2020})},\ \Eprint
  {https://arxiv.org/abs/2003.07384} {arXiv:2003.07384 [gr-qc]} \BibitemShut
  {NoStop}%
\bibitem [{\citenamefont {Beltr{\'a}n~Jim{\'e}nez}\ and\ \citenamefont
  {Maldonado~Torralba}(2020)}]{BeltranJimenez:2019hrm}%
  \BibitemOpen
  \bibfield  {author} {\bibinfo {author} {\bibfnamefont {J.}~\bibnamefont
  {Beltr{\'a}n~Jim{\'e}nez}}\ and\ \bibinfo {author} {\bibfnamefont {F.~J.}\
  \bibnamefont {Maldonado~Torralba}},\ }\bibfield  {title} {\bibinfo {title}
  {{Revisiting the stability of quadratic Poincar{\'e} gauge gravity}},\ }\href
  {https://doi.org/10.1140/epjc/s10052-020-8163-8} {\bibfield  {journal}
  {\bibinfo  {journal} {Eur. Phys. J. C}\ }\textbf {\bibinfo {volume} {80}},\
  \bibinfo {pages} {611} (\bibinfo {year} {2020})},\ \Eprint
  {https://arxiv.org/abs/1910.07506} {arXiv:1910.07506 [gr-qc]} \BibitemShut
  {NoStop}%
\bibitem [{\citenamefont {Iosifidis}\ and\ \citenamefont
  {Koivisto}(2024)}]{Iosifidis:2023kyf}%
  \BibitemOpen
  \bibfield  {author} {\bibinfo {author} {\bibfnamefont {D.}~\bibnamefont
  {Iosifidis}}\ and\ \bibinfo {author} {\bibfnamefont {T.~S.}\ \bibnamefont
  {Koivisto}},\ }\bibfield  {title} {\bibinfo {title} {{Hyperhydrodynamics:
  relativistic viscous fluids from hypermomentum}},\ }\href
  {https://doi.org/10.1088/1475-7516/2024/05/001} {\bibfield  {journal}
  {\bibinfo  {journal} {JCAP}\ }\textbf {\bibinfo {volume} {05}}\bibfield
  {number} {\bibinfo  {number} { (2024)},\ \bibinfo {pages} {001}},\ }\Eprint
  {https://arxiv.org/abs/2312.06780} {arXiv:2312.06780 [gr-qc]} \BibitemShut
  {NoStop}%
\bibitem [{\citenamefont {Janssen}\ \emph {et~al.}(2019)\citenamefont
  {Janssen}, \citenamefont {Jim{\'e}nez-Cano},\ and\ \citenamefont
  {Orejuela}}]{Janssen:2019doc}%
  \BibitemOpen
  \bibfield  {author} {\bibinfo {author} {\bibfnamefont {B.}~\bibnamefont
  {Janssen}}, \bibinfo {author} {\bibfnamefont {A.}~\bibnamefont
  {Jim{\'e}nez-Cano}},\ and\ \bibinfo {author} {\bibfnamefont {J.~A.}\
  \bibnamefont {Orejuela}},\ }\bibfield  {title} {\bibinfo {title} {{A
  non-trivial connection for the metric-affine Gauss-Bonnet theory in $D =
  4$}},\ }\href {https://doi.org/10.1016/j.physletb.2019.06.002} {\bibfield
  {journal} {\bibinfo  {journal} {Phys. Lett. B}\ }\textbf {\bibinfo {volume}
  {795}},\ \bibinfo {pages} {42} (\bibinfo {year} {2019})},\ \Eprint
  {https://arxiv.org/abs/1903.00280} {arXiv:1903.00280 [gr-qc]} \BibitemShut
  {NoStop}%
\bibitem [{\citenamefont {Janssen}\ and\ \citenamefont
  {Jim{\'e}nez-Cano}(2019)}]{Janssen:2019uao}%
  \BibitemOpen
  \bibfield  {author} {\bibinfo {author} {\bibfnamefont {B.}~\bibnamefont
  {Janssen}}\ and\ \bibinfo {author} {\bibfnamefont {A.}~\bibnamefont
  {Jim{\'e}nez-Cano}},\ }\bibfield  {title} {\bibinfo {title} {{On the
  topological character of metric-affine Lovelock Lagrangians in critical
  dimensions}},\ }\href {https://doi.org/10.1016/j.physletb.2019.134996}
  {\bibfield  {journal} {\bibinfo  {journal} {Phys. Lett. B}\ }\textbf
  {\bibinfo {volume} {798}},\ \bibinfo {pages} {134996} (\bibinfo {year}
  {2019})},\ \Eprint {https://arxiv.org/abs/1907.12100} {arXiv:1907.12100
  [gr-qc]} \BibitemShut {NoStop}%
\bibitem [{\citenamefont {Baker}\ \emph {et~al.}(2017)\citenamefont {Baker}
  \emph {et~al.}}]{Baker:2017hug}%
  \BibitemOpen
  \bibfield  {author} {\bibinfo {author} {\bibfnamefont {T.}~\bibnamefont
  {Baker}} \emph {et~al.},\ }\bibfield  {title} {\bibinfo {title} {{Strong
  constraints on cosmological gravity from GW170817 and GRB 170817A}},\ }\href
  {https://doi.org/10.1103/PhysRevLett.119.251301} {\bibfield  {journal}
  {\bibinfo  {journal} {Phys. Rev. Lett.}\ }\textbf {\bibinfo {volume} {119}},\
  \bibinfo {pages} {251301} (\bibinfo {year} {2017})},\ \Eprint
  {https://arxiv.org/abs/1710.06394} {arXiv:1710.06394 [astro-ph.CO]}
  \BibitemShut {NoStop}%
\bibitem [{\citenamefont {Creminelli}\ and\ \citenamefont
  {Vernizzi}(2017)}]{Creminelli:2017sry}%
  \BibitemOpen
  \bibfield  {author} {\bibinfo {author} {\bibfnamefont {P.}~\bibnamefont
  {Creminelli}}\ and\ \bibinfo {author} {\bibfnamefont {F.}~\bibnamefont
  {Vernizzi}},\ }\bibfield  {title} {\bibinfo {title} {{Dark Energy after
  GW170817 and GRB170817A}},\ }\href
  {https://doi.org/10.1103/PhysRevLett.119.251302} {\bibfield  {journal}
  {\bibinfo  {journal} {Phys. Rev. Lett.}\ }\textbf {\bibinfo {volume} {119}},\
  \bibinfo {pages} {251302} (\bibinfo {year} {2017})},\ \Eprint
  {https://arxiv.org/abs/1710.05877} {arXiv:1710.05877 [astro-ph.CO]}
  \BibitemShut {NoStop}%
\bibitem [{\citenamefont {Sakstein}\ and\ \citenamefont
  {Jain}(2017)}]{Sakstein:2017xjx}%
  \BibitemOpen
  \bibfield  {author} {\bibinfo {author} {\bibfnamefont {J.}~\bibnamefont
  {Sakstein}}\ and\ \bibinfo {author} {\bibfnamefont {B.}~\bibnamefont
  {Jain}},\ }\bibfield  {title} {\bibinfo {title} {{Implications of the Neutron
  Star Merger GW170817 for Cosmological Scalar-Tensor Theories}},\ }\href
  {https://doi.org/10.1103/PhysRevLett.119.251303} {\bibfield  {journal}
  {\bibinfo  {journal} {Phys. Rev. Lett.}\ }\textbf {\bibinfo {volume} {119}},\
  \bibinfo {pages} {251303} (\bibinfo {year} {2017})},\ \Eprint
  {https://arxiv.org/abs/1710.05893} {arXiv:1710.05893 [astro-ph.CO]}
  \BibitemShut {NoStop}%
\bibitem [{\citenamefont {Ezquiaga}\ and\ \citenamefont
  {Zumalac{\'a}rregui}(2017)}]{Ezquiaga:2017ekz}%
  \BibitemOpen
  \bibfield  {author} {\bibinfo {author} {\bibfnamefont {J.~M.}\ \bibnamefont
  {Ezquiaga}}\ and\ \bibinfo {author} {\bibfnamefont {M.}~\bibnamefont
  {Zumalac{\'a}rregui}},\ }\bibfield  {title} {\bibinfo {title} {{Dark Energy
  After GW170817: Dead Ends and the Road Ahead}},\ }\href
  {https://doi.org/10.1103/PhysRevLett.119.251304} {\bibfield  {journal}
  {\bibinfo  {journal} {Phys. Rev. Lett.}\ }\textbf {\bibinfo {volume} {119}},\
  \bibinfo {pages} {251304} (\bibinfo {year} {2017})},\ \Eprint
  {https://arxiv.org/abs/1710.05901} {arXiv:1710.05901 [astro-ph.CO]}
  \BibitemShut {NoStop}%
\bibitem [{\citenamefont {Callister}\ \emph {et~al.}(2017)\citenamefont
  {Callister} \emph {et~al.}}]{Callister:2017ocg}%
  \BibitemOpen
  \bibfield  {author} {\bibinfo {author} {\bibfnamefont {T.}~\bibnamefont
  {Callister}} \emph {et~al.},\ }\bibfield  {title} {\bibinfo {title}
  {{Polarization-based Tests of Gravity with the Stochastic Gravitational-Wave
  Background}},\ }\href {https://doi.org/10.1103/PhysRevX.7.041058} {\bibfield
  {journal} {\bibinfo  {journal} {Phys. Rev. X}\ }\textbf {\bibinfo {volume}
  {7}},\ \bibinfo {pages} {041058} (\bibinfo {year} {2017})},\ \Eprint
  {https://arxiv.org/abs/1704.08373} {arXiv:1704.08373 [gr-qc]} \BibitemShut
  {NoStop}%
\bibitem [{\citenamefont {Abbott}\ \emph {et~al.}(2019)\citenamefont {Abbott}
  \emph {et~al.}}]{LIGOScientific:2018dkp}%
  \BibitemOpen
  \bibfield  {author} {\bibinfo {author} {\bibfnamefont {B.~P.}\ \bibnamefont
  {Abbott}} \emph {et~al.} (\bibinfo {collaboration} {LIGO Scientific,
  Virgo}),\ }\bibfield  {title} {\bibinfo {title} {{Tests of General Relativity
  with GW170817}},\ }\href {https://doi.org/10.1103/PhysRevLett.123.011102}
  {\bibfield  {journal} {\bibinfo  {journal} {Phys. Rev. Lett.}\ }\textbf
  {\bibinfo {volume} {123}},\ \bibinfo {pages} {011102} (\bibinfo {year}
  {2019})},\ \Eprint {https://arxiv.org/abs/1811.00364} {arXiv:1811.00364
  [gr-qc]} \BibitemShut {NoStop}%
\bibitem [{\citenamefont {Hagihara}\ \emph {et~al.}(2018)\citenamefont
  {Hagihara}, \citenamefont {Era}, \citenamefont {Iikawa},\ and\ \citenamefont
  {Asada}}]{Hagihara:2018azu}%
  \BibitemOpen
  \bibfield  {author} {\bibinfo {author} {\bibfnamefont {Y.}~\bibnamefont
  {Hagihara}}, \bibinfo {author} {\bibfnamefont {N.}~\bibnamefont {Era}},
  \bibinfo {author} {\bibfnamefont {D.}~\bibnamefont {Iikawa}},\ and\ \bibinfo
  {author} {\bibfnamefont {H.}~\bibnamefont {Asada}},\ }\bibfield  {title}
  {\bibinfo {title} {{Probing gravitational wave polarizations with Advanced
  LIGO, Advanced Virgo and KAGRA}},\ }\href
  {https://doi.org/10.1103/PhysRevD.98.064035} {\bibfield  {journal} {\bibinfo
  {journal} {Phys. Rev. D}\ }\textbf {\bibinfo {volume} {98}},\ \bibinfo
  {pages} {064035} (\bibinfo {year} {2018})},\ \Eprint
  {https://arxiv.org/abs/1807.07234} {arXiv:1807.07234 [gr-qc]} \BibitemShut
  {NoStop}%
\bibitem [{\citenamefont {Abbott}\ \emph {et~al.}(2021)\citenamefont {Abbott}
  \emph {et~al.}}]{KAGRA:2021kbb}%
  \BibitemOpen
  \bibfield  {author} {\bibinfo {author} {\bibfnamefont {R.}~\bibnamefont
  {Abbott}} \emph {et~al.} (\bibinfo {collaboration} {KAGRA, Virgo, LIGO
  Scientific}),\ }\bibfield  {title} {\bibinfo {title} {{Upper limits on the
  isotropic gravitational-wave background from Advanced LIGO and Advanced
  Virgo{\textquoteright}s third observing run}},\ }\href
  {https://doi.org/10.1103/PhysRevD.104.022004} {\bibfield  {journal} {\bibinfo
   {journal} {Phys. Rev. D}\ }\textbf {\bibinfo {volume} {104}},\ \bibinfo
  {pages} {022004} (\bibinfo {year} {2021})},\ \Eprint
  {https://arxiv.org/abs/2101.12130} {arXiv:2101.12130 [gr-qc]} \BibitemShut
  {NoStop}%
\bibitem [{\citenamefont {Yunes}\ and\ \citenamefont
  {Hughes}(2010)}]{Yunes:2010qb}%
  \BibitemOpen
  \bibfield  {author} {\bibinfo {author} {\bibfnamefont {N.}~\bibnamefont
  {Yunes}}\ and\ \bibinfo {author} {\bibfnamefont {S.~A.}\ \bibnamefont
  {Hughes}},\ }\bibfield  {title} {\bibinfo {title} {{Binary Pulsar Constraints
  on the Parameterized post-Einsteinian Framework}},\ }\href
  {https://doi.org/10.1103/PhysRevD.82.082002} {\bibfield  {journal} {\bibinfo
  {journal} {Phys. Rev. D}\ }\textbf {\bibinfo {volume} {82}},\ \bibinfo
  {pages} {082002} (\bibinfo {year} {2010})},\ \Eprint
  {https://arxiv.org/abs/1007.1995} {arXiv:1007.1995 [gr-qc]} \BibitemShut
  {NoStop}%
\bibitem [{\citenamefont {Zhu}\ \emph {et~al.}(2019)\citenamefont {Zhu} \emph
  {et~al.}}]{Zhu:2018etc}%
  \BibitemOpen
  \bibfield  {author} {\bibinfo {author} {\bibfnamefont {W.~W.}\ \bibnamefont
  {Zhu}} \emph {et~al.},\ }\bibfield  {title} {\bibinfo {title} {{Tests of
  Gravitational Symmetries with Pulsar Binary J1713+0747}},\ }\href
  {https://doi.org/10.1093/mnras/sty2905} {\bibfield  {journal} {\bibinfo
  {journal} {Mon. Not. Roy. Astron. Soc.}\ }\textbf {\bibinfo {volume} {482}},\
  \bibinfo {pages} {3249} (\bibinfo {year} {2019})},\ \Eprint
  {https://arxiv.org/abs/1802.09206} {arXiv:1802.09206 [astro-ph.HE]}
  \BibitemShut {NoStop}%
\bibitem [{\citenamefont {Callen}(1985)}]{callen}%
  \BibitemOpen
  \bibfield  {author} {\bibinfo {author} {\bibfnamefont {H.~B.}\ \bibnamefont
  {Callen}},\ }\href
  {https://www.wiley.com/Thermodynamics+and+an+Introduction+to+Thermostatistics%2C+2nd+Edition-p-9780471862567}
  {\emph {\bibinfo {title} {Thermodynamics and an Introduction to
  Thermostatistics}}}\ (\bibinfo  {publisher} {John Wiley \& Sons},\ \bibinfo
  {year} {1985})\BibitemShut {NoStop}%
\bibitem [{\citenamefont {Wald}(1984)}]{wald2010general}%
  \BibitemOpen
  \bibfield  {author} {\bibinfo {author} {\bibfnamefont {R.~M.}\ \bibnamefont
  {Wald}},\ }\href
  {https://press.uchicago.edu/ucp/books/book/chicago/G/bo5952261.html} {\emph
  {\bibinfo {title} {General Relativity}}}\ (\bibinfo  {publisher} {The
  University of Chicago Press},\ \bibinfo {year} {1984})\BibitemShut {NoStop}%
\bibitem [{\citenamefont {Guarnizo}\ \emph {et~al.}(2010)\citenamefont
  {Guarnizo}, \citenamefont {Castañeda},\ and\ \citenamefont
  {Tejeiro}}]{Guarnizo:2010xr}%
  \BibitemOpen
  \bibfield  {author} {\bibinfo {author} {\bibfnamefont {A.}~\bibnamefont
  {Guarnizo}}, \bibinfo {author} {\bibfnamefont {L.}~\bibnamefont
  {Castañeda}},\ and\ \bibinfo {author} {\bibfnamefont {J.~M.}\ \bibnamefont
  {Tejeiro}},\ }\bibfield  {title} {\bibinfo {title} {{Boundary Term in Metric
  $f(R)$ Gravity: Field Equations in the Metric Formalism}},\ }\href
  {https://doi.org/10.1007/s10714-010-1012-6} {\bibfield  {journal} {\bibinfo
  {journal} {Gen. Rel. Grav.}\ }\textbf {\bibinfo {volume} {42}},\ \bibinfo
  {pages} {2713} (\bibinfo {year} {2010})},\ \Eprint
  {https://arxiv.org/abs/1002.0617} {arXiv:1002.0617 [gr-qc]} \BibitemShut
  {NoStop}%
\bibitem [{\citenamefont {Bravo~Medina}\ \emph {et~al.}(2019)\citenamefont
  {Bravo~Medina}, \citenamefont {Nowakowski},\ and\ \citenamefont
  {Batic}}]{Medina:2018rnl}%
  \BibitemOpen
  \bibfield  {author} {\bibinfo {author} {\bibfnamefont {S.}~\bibnamefont
  {Bravo~Medina}}, \bibinfo {author} {\bibfnamefont {M.}~\bibnamefont
  {Nowakowski}},\ and\ \bibinfo {author} {\bibfnamefont {D.}~\bibnamefont
  {Batic}},\ }\bibfield  {title} {\bibinfo {title} {{Einstein-Cartan
  Cosmologies}},\ }\href {https://doi.org/10.1016/j.aop.2018.11.002} {\bibfield
   {journal} {\bibinfo  {journal} {Annals Phys.}\ }\textbf {\bibinfo {volume}
  {400}},\ \bibinfo {pages} {64} (\bibinfo {year} {2019})},\ \Eprint
  {https://arxiv.org/abs/1812.04589} {arXiv:1812.04589 [gr-qc]} \BibitemShut
  {NoStop}%
\bibitem [{\citenamefont {Ort{\'\i}n}(2004)}]{ortin2004gravity}%
  \BibitemOpen
  \bibfield  {author} {\bibinfo {author} {\bibfnamefont {T.}~\bibnamefont
  {Ort{\'\i}n}},\ }\href {https://doi.org/10.1017/CBO9780511616563} {\emph
  {\bibinfo {title} {Gravity and Strings}}},\ Cambridge Monographs on
  Mathematical Physics\ (\bibinfo  {publisher} {Cambridge University Press},\
  \bibinfo {year} {2004})\BibitemShut {NoStop}%
\end{thebibliography}%
\end{document}